\documentclass[aps,pre,preprint,groupedaddress]{revtex4-1}
\usepackage{comment}
\usepackage{amsmath, amssymb} 
\usepackage{mathtools}

\usepackage[dvipdfmx]{color}
\usepackage[dvipdfmx]{graphicx}

\usepackage{array}

\begin{document}

%\preprint{}

\title{A new Granger causality measure for eliminating the confounding influence of latent common inputs}


\author{Takashi Arai}
%\thanks{}
\affiliation{The Center for Data Science Education and Research, Shiga University, Hikone 522-8522, Japan}


%\date{\today}

\begin{abstract}
In this paper, we propose a new Granger causality measure which is robust against the confounding influence of latent common inputs.
This measure is inspired by partial Granger causality in the literature, and its variant.
Using numerical experiments we first show that the test statistics for detecting directed interactions between time series approximately obey the $F$-distributions when there are no interactions.
Then, we propose a practical procedure for inferring directed interactions, which is based on the idea of multiple statistical test in situations where the confounding influence of latent common inputs may exist.
The results of numerical experiments demonstrate that the proposed method successfully eliminates the influence of latent common inputs while the normal Granger causality method detects spurious interactions due to the influence of the confounder.
\end{abstract}

%\keywords{} \keywords{stochastic process \and su(2) algebra \and spin coherent state \and path integral}

\maketitle

%\section{}

\section{Introduction}
The Granger causality test is a method for inferring causal interactions between time series by a statistical test.
This method is based on the idea by Wiener that if past information of one time series improves the prediction of another time series, there is a causal influence~\cite{Wiener_1956}.
Granger formalized this notion of causality in the framework on autoregressive (AR) modeling of multivariate time series in the econometrics literature~\cite{Granger_1969}.
Strictly speaking, Granger causality does not mean causality in the true sense, although it should be considered as a necessary condition for the true causality.
However, the method has gained popularity in a wide range of fields, such as econometrics and biology due to its conceptual simplicity and ease of implementation~\cite{Granger_1980,Sameshima_1998,Gourevitch_2006}.
For example, in the field of biology, the Granger causality test was applied to electroencephalogram~\cite{EEG_2012}, functional magnetic resonance imaging of Blood Oxygen Level Dependent signals~\cite{appl_book,Chen_2006,wem,Seth_jns}, multi-electrode arrays of local field potentials and Calcium imaging of neuronal activity~\cite{CID_2015}.
Also, more recently, Granger causality has extended to point process time series and applied to spike activity of neurons~\cite{Stevenson_2009,Kim_2011,Eden_2013}.

Nowadays, application of the Granger causality test is more common.
However, when we apply the Granger causality test to real data, we must pay attention to confounding environmental influences.
In fact, in many experimental settings, we are only able to record a subset of all related variables in a system.
In such a case, applications of Granger causality may detect spurious directed interactions due to the influence of latent common inputs and unobserved variables.
For example, one time series may falsely appear to cause another if they are both influenced by a third time series or a common input but with a delay.
In addition, data analysis in practice involves the step of model selection, in which a relevant set of variables is selected for analysis~\cite{brain_selection}.
Then, this step is likely to exclude some relevant variables, which can lead to the detection of apparent causal interactions that are actually spurious~\cite{Pearl_2000}.
Hence, controlling for latent common inputs and latent variables is a critical issue when applying Granger causality to experimental data.

Partial Granger causality was developed as a method for eliminating the confounding influence of latent common inputs and unobserved variables~\cite{Guo_2008}.
According to what the authors in the literature say, the method was inspired by the partial correlation in statistics.
They insisted that partial Granger causality does estimate directed interactions more robustly than conditional Granger causality against the influence of latent common inputs and latent variables.
In the recent literature we can find some reviews and applications of this type of Granger causality~\cite{Smith_2011}.
Although the idea of partial Granger causality is fascinating and the method is applied to various systems, the justification of the method is arguable~\cite{dpgc}.
In the original paper, it was insisted that the sampling distribution of the test statistics $F_1$ cannot be determined analytically, and one has to resort to the bootstrap technique.
Furthermore, the test statistics can take a negative value, which was reasoned as being subtracted by an additional term.
By contrast, in the subsequent study in which partial Granger causality was extended to multivariate system, it was argued that the test statistics cannot take a negative value~\cite{Barrett_2010}.
In another literature, it was argued that the negative value in the test statistics of partial Granger causality is a serious flaw, and this undermines the credibility of the obtained results and, thus, the validity of the approach~\cite{dpgc}.
Therefore, it remains unclear that at which circumstance partial Granger causality is available and not even in numerical experiments.

In this paper, modifying the idea of partial Granger causality, we propose a new Granger causality measure which is robust against the confounding influence of latent common inputs.
We first point out that the original partial Granger causality is attempting to eliminate the influence of the confounder between the target variable and conditioning variable, rather than the target variable and source variable.
Then, we introduce a Granger causality measure by eliminating the influence of latent common inputs between the target variable and source variable. 
We also discuss the way of parameter estimation for the model.
In our parameter estimation procedure, the test statistics cannot take a negative value.
Of the influence of latent confounder, this paper only considers the latent common inputs, and the latent variables are not considered which has a worse effect on the estimation of directed interactions~\cite{dpgc}.
The reason for this is that the influence of latent variables have to be dealt with in the framework of the Moving Average (MA) model, rather than the AR model.

This paper is organized as follows.
In Sec.~\ref{sec: proposed_method}, we review partial Granger causality and introduce our Granger causality measure to illustrate and highlight the difference between partial Granger causality and proposed Granger causality introduced in this paper.
Furthermore, we discuss the distribution that the test statistics obeys by using the likelihood function of the observed time series.
We also discuss the way of parameter estimation for the model and the reason why the test statistics can take a negative value in the literature.
In Sec.~\ref{sec: numerical_experiment}, we show the validity of our measure using the data generated by the models, which are variants of the models extensively studied in the literature for numerical experiments.
Furthermore, we propose a practical procedure for estimating directed interactions where latent common inputs may exist.
Sec.~\ref{sec: conclusion} is devoted to conclusions.

\section{Proposed measure for Granger causality \label{sec: proposed_method}}
In this section, we review partial Granger causality to illustrate and highlight the difference between partial Granger causality and proposed Granger causality introduced in this paper, and point out the issue of partial Granger causality.
Then, we introduce our new Granger causality measure.

\subsection{Partial Granger causality}
Seth et al., proposed the following idea of Granger causality which eliminates the influence of latent confounder.
In their notation, they considered the following AR model for the null hypothesis:
\begin{align}
&X_t = \sum_{i=1}^{\infty} a_{1 i} X_{t-i} + \sum_{i=1}^{\infty} c_{1 i} Z_{t-i} + u_{1 t}, \\
&Z_t = \sum_{i=1}^{\infty} b_{1 i} Z_{t-i} + \sum_{i=1}^{\infty} d_{1 i} X_{t-i} + u_{2 t},
\end{align}
where $u_{1 t}$, $u_{2 t}$ are noise terms which have zero mean and involve the influence of latent common inputs and latent variables.
They incorporated the influence into the model explicitly by the following covariance matrix:
\begin{align}
S =& \begin{pmatrix} S_{11} & S_{12} \\ S_{21} & S_{22} \end{pmatrix} \notag \\
=& \begin{pmatrix} \mathrm{var}(u_{1 t}) & \mathrm{cov}(u_{1 t}, u_{2 t}) \\ \mathrm{cov}(u_{2 t}, u_{1 t}) & \mathrm{var}(u_{2 t}) \end{pmatrix}.
\end{align}
In the same way, the model with directed interactions for the alternative hypothesis is expressed as follows:
\begin{align}
&X_t = \sum_{i=1}^{\infty} a_{2 i} X_{t-i} + \sum_{i=1}^{\infty} b_{2 i} Y_{t-i} + \sum_{i=1}^{\infty} c_{2 i} Z_{t-i} + u_{3 t}, \\
&Y_t = \sum_{i=1}^{\infty} d_{2 i} X_{t-i} + \sum_{i=1}^{\infty} e_{2 i} Y_{t-i} + \sum_{i=1}^{\infty} f_{2 i} Z_{t-i} + u_{4 t}, \\
&Z_t = \sum_{i=1}^{\infty} g_{2 i} X_{t-i} + \sum_{i=1}^{\infty} h_{2 i} Y_{t-i} + \sum_{i=1}^{\infty} k_{2 i} Z_{t-i} + u_{5 t},
\end{align}
where each noise has a following covariance matrix:
\begin{align}
\Sigma =& \begin{pmatrix} \Sigma_{11} & \Sigma_{12} & \Sigma_{13} \\ \Sigma_{21} & \Sigma_{22} & \Sigma_{23} \\ \Sigma_{31} & \Sigma_{32} & \Sigma_{33}  \end{pmatrix} \notag \\
=& \begin{pmatrix} \mathrm{var}(u_{3 t}) & \mathrm{cov}(u_{3 t}, u_{4 t}) & \mathrm{cov}(u_{3 t}, u_{5 t}) \\
                                    \mathrm{cov}(u_{4 t}, u_{3 t})         & \mathrm{var}(u_{4 t})  & \mathrm{cov}(u_{4 t}, u_{5 t}) \\
                                    \mathrm{cov}(u_{5 t}, u_{3 t})         & \mathrm{cov}(u_{5 t}, u_{4 t})  & \mathrm{var}(u_{5 t}) \end{pmatrix}.
\end{align}
Then, for each model, they introduce the variance of noise for the target variable $X_t$ by eliminating the influence of the noise for $Z_t$ as
\begin{align}
R_{XX | Z}^{(1)}  \equiv & S_{11} - S_{12} S_{22}^{-1} S_{21}, \label{eq: R_1} \\
R_{XX | Z}^{(2)}  \equiv & \Sigma_{11} - \Sigma_{13} \Sigma_{33}^{-1} \Sigma_{31}. \label{eq: R_2}
\end{align}
Since $R_{XX|Z}^{(1)}$, $R_{XX|Z}^{(2)}$ are variance of noise for target variable with the influence of latent common inputs and latent variables eliminated, they insisted that directed interactions under the existence of latent common inputs and latent variables can be estimated by the analysis of variance of $R_{XX|Z}^{(1)}$, $R_{XX|Z}^{(2)}$.
The test statistic is given by
\begin{align}
F_1 = \ln \Biggl( \frac{ R_{XX | Z}^{(1)}  }{ R_{XX | Z}^{(2)}  } \Biggr).
\end{align}
It is because the test statistic $F_1$ quantifies the decrease of variance by including interaction terms with the influence of confounder eliminated.
They called the above statistical test as partial Granger causality since it was inspired by the partial correlation in statistics.

\subsection{Proposed statistical measure}
In our opinion, partial Granger causality introduced in the previous subsection has a conceptual issue.
The variance, Eqs.~(\ref{eq: R_1}) and (\ref{eq: R_2}), eliminates the noise correlation between the target variable $X_t$ and the conditioning variable $Z_t$.
That is, they eliminate the confounding environmental influence between the target variable and the conditioning variable.
However, when we are going to estimate directed interactions from the source variable $Y_t$ to the target variable $X_t$, the influence that should be removed is that between the source variable and the target variable, which can cause a spurious interaction from $Y_t$ to $X_t$.
Therefore, we define here the variance by eliminating the influence of latent common inputs between the source variable and the target variable in a simplest example of the AR model.
Then, using this variance, we propose a new Granger causality test for directed interactions under the existence of latent common inputs.

The model we suppose consists of two time series of the following AR models $x_t$, $y_t$, however, the noise structure is different from the usual AR model: 
\begin{align}
x_t =& a x_{t-1} + c y_{t-1} + \varepsilon_t, \label{eq: naive_AR_x} \\
y_t =& b y_{t-1} + \xi_{t}. \label{eq: naive_AR_y}
\end{align}
Here, the noise term $\varepsilon_t$ and $\xi_t$ are zero mean Gaussian noises, although they have a different-time correlation due to the latent common inputs with delay.
That is, the covariance matrix of the noise is expressed as follows,
\begin{align}
\Sigma =& \begin{pmatrix} \Sigma_{xx} & \Sigma_{xy} \\ \Sigma_{yx} & \Sigma_{yy} \end{pmatrix} 
\equiv \begin{pmatrix} \mathrm{var}(\varepsilon_t) & \mathrm{cov}(\varepsilon_t, \xi_{t-1}) \\ \mathrm{cov}(\xi_{t-1}, \varepsilon_t) & \mathrm{var}(\xi_{t-1}) \end{pmatrix} \notag \\
= & \begin{pmatrix} \sigma_x^2 & \rho \sigma_x \sigma_y \\ \rho \sigma_x \sigma_y & \sigma_y^2 \end{pmatrix},
\end{align}
where $\rho$ is a correlation coefficient between $\varepsilon_t$ and $\xi_{t-1}$, which means that the latent common input reaches $x_t$ with delay compared to $y_t$.
These noises do not have autocorrelation, $\mathrm{cov}(\varepsilon_t, \varepsilon_{t-i}) = \mathrm{cov}(\xi_t, \xi_{t-i}) = 0, \hspace{0.2cm} i \in \mathbb{N}$.
In this paper, we express the influence of latent variables by an auto-correlated noise.
Our interest is the estimation of interactions from $y_t$ to $x_t$, that is, whether the model parameter $c$ exists or not.
For simplicity, we assume that the interaction of opposite direction, from $x_t$ to $y_t$, does not exist, that is, the system of our interest comprises asymmetric network structure.

Here, we note the following fact.
In the original paper of partial Granger causality, two types of noise term which are generated by the confounding environmental influence are considered.
The first is generated by latent common inputs, which has a cross correlation between each time series but does not have autocorrelation.
In addition, only the equal-time cross-correlation was considered.
However, the equal-time correlated noise generated by latent common inputs does not produce a spurious directed interaction, this fact will be numerically confirmed later.
Latent common inputs with delays could generate a spurious directed interaction when one uses the normal Granger causality test.
The influence of latent common inputs can be removed by the method proposed in this paper.
The second is the noise generated by the latent variables.
The influence of latent variables is expressed by an auto-correlated noise, $\mathrm{cov}(\varepsilon_t, \varepsilon_{t-i}) \neq 0, \hspace{0.2cm} i \in \mathbb{N}$.
However, strictly speaking, an auto-correlated noise is beyond the scope of the AR model.
Therefore, in this paper, we do not suppose the existence of latent variables.
The influence of latent variables requires Granger causality constructed by the MA model.
The study in this direction has been worked on in the framework of the state space model~\cite{ssgc}.

To construct our new measure for Granger causality, it is useful to reparametrize the noise parameter $(\sigma_x, \sigma_y, \rho)$ to $(\tau, \sigma_y, \eta)$ as
\begin{align}
\Sigma = \begin{pmatrix} \tau^2 + \eta^2 \sigma_y^2 & \eta \sigma_y^2 \\ \eta \sigma_y^2 & \sigma_y^2 \end{pmatrix},
\end{align}
where
\begin{align}
\tau^2 = \Sigma_{xx} - \Sigma_{xy} \Sigma_{yy}^{-1} \Sigma_{yx} \label{eq: tau}
\end{align}
is the variance of the noise with the influence of latent common inputs eliminated similar to partial Granger causality.
This reparameterization is equivalent to following linear relationship between $\varepsilon_t$ and $\xi_t$:
\begin{align}
\varepsilon_t = \eta \xi_{t-1} + \omega_t,
\end{align}
where the noise $\omega_t$ has following property:
\begin{align}
\omega_t \sim  \mathcal{N}(0, \tau^2),
\end{align}
\begin{align}
 \mathrm{cov}(\omega_t, \xi_{t-1} ) 
= \mathrm{cov}(\omega_t, \varepsilon_t ) = 0.
\end{align}

Using the above reparameterization, the model, Eqs.~(\ref{eq: naive_AR_x}) and (\ref{eq: naive_AR_y}), can be re-expressed as
\begin{align}
x_t =& a x_{t-1} + c y_{t-1} + \eta \xi_{t-1} + \omega_t, \label{eq: AR_x} \\
y_t =& b y_{t-1} + \xi_{t}. \label{eq: AR_y}
\end{align}
Since the conditional variance, $\tau^2 = \Sigma_{xx} - \Sigma_{xy} \Sigma_{yy}^{-1} \Sigma_{yx}$, is a variance of noise for $x_t$ from which the correlated component with the noise for $y_t$ was eliminated, we propose a statistical measure for Granger causality in detecting directed interactions as whether the variance of $\omega_t$ decreases significantly by including the interaction parameter $c$.
That is, we set the models of the null hypothesis and the alternative hypothesis as follows:
\begin{align}
\begin{cases}
H_0 :& \hspace{0.02cm} c=0, \\
H_1 :& \hspace{0.02cm} c\neq 0.
\end{cases}
\end{align}

\subsection{Interpretation by the likelihood function}
In this subsection, we re-express the formulation explained in the previous subsection in terms of the likelihood function in order to consider the distribution that our test statistic obeys.
First, we note that the variance of the noise, Eq.~(\ref{eq: tau}), is formally equivalent to that of the conditional Gaussian distribution.
In fact, the probability of the AR model, Eq.~(\ref{eq: AR_x}), can be expressed as a conditional Gaussian distribution:
\begin{align}
p(x_t| x_{t-1}, y_{t-1}, y_{t-2}) = \mathcal{N}(\mu_t^{(1)}, \Sigma_{x|y}),
\end{align}
where
\begin{align}
\mu_t^{(1)} =& a x_{t-1} + c y_{t-1} +  \eta ( y_{t-1} - b y_{t-2} ) \notag \\
=& a x_{t-1} + c y_{t-1} + \eta \xi_{t-1}, \\
\Sigma_{x|y} =&  \Sigma_{xx} - \Sigma_{xy}\Sigma_{yy}^{-1} \Sigma_{yx} \notag \\
 =& \tau^2.
\end{align}
Therefore, extracting the correlated component of noise due to latent common inputs is equivalent to conditioning on $y_t$.

Now, we consider the situation where time series $x_t$ and $y_t$ are observed, $\mathbf{x} = (x_T, x_{T-1}, \dots, x_2)$, $\mathbf{y} = (y_{T-1}, y_{T-2}, \dots, y_2, y_1)$.
We denote the model parameters as $\boldsymbol{\beta} = (a, c, \eta)$ and $b$.
Then, the overall likelihood of the time series except for initial values is given by the following joint probability distribution:
\begin{align}
L(\boldsymbol{\beta}, b | \mathbf{x}, \mathbf{y}) = p(x_T, x_{T-1}, \dots, x_3, y_{T-1}, \dots, y_1 | x_2, y_1, \boldsymbol{\beta}, b).
\end{align}
Since the time series is modeled as the AR model and the network structure of the time series is assumed to be asymmetric, the above likelihood function can be divided into the following product of two conditional distributions:
\begin{align}
L(\boldsymbol{\beta}, b | \mathbf{x}, \mathbf{y})  =& p(x_T, x_{T-1}, \dots, y_{T-1}, y_{T-2}, \dots | x_2, y_1, \boldsymbol{\beta}, b)  \notag \\
=& p(x_T | x_{T-1} y_{T-1}, y_{T-2}  | x_2, y_1, \boldsymbol{\beta}, b ) \ p(x_{T-1}, x_{T-2}, \dots, y_{T-1}, y_{T-2}, \dots  | x_2, y_1, \boldsymbol{\beta}, b)  \notag \\
& \vdots \notag \\
=& \prod_{t=3}^T p(x_t | x_{t-1}, y_{t-1}, y_{t-2} , \boldsymbol{\beta}, b) \prod_{t=2}^{T-1} p(y_t | y_{t-1} ,b ) \notag \\
\equiv & L_{x|y}(\boldsymbol{\beta}, b | \mathbf{x}, \mathbf{y} ) \ L_y(b | \mathbf{y}).
\end{align}
Therefore, testing whether the variance $\Sigma_{x|y}$ decreases significantly, by including directed interactions, is equivalent to test whether the conditional likelihood $L_{x|y}$ increases significantly.

\subsection{Parameter estimation}
Log-likelihood functions of the observed time series are given by
\begin{align}
\log L_{x|y}( \boldsymbol{\beta}, b | \mathbf{x}, \mathbf{y} ) 
 =& -\frac{T-2}{2} \log \tau^2 - \frac{1}{2 \tau^2} \sum_{t=3}^{T} \Bigl[ x_t - a x_{t-1} - c y_{t-1} - \eta (y_{t-1} - b y_{t-2} ) \Bigr]^2 \notag \\
=& -\frac{T-2}{2} \log \tau^2 - \frac{1}{2 \tau^2} \sum_{t=3}^{T} \Bigl[ x_t - a x_{t-1} - c y_{t-1} - \eta \xi_{t-1} \Bigr]^2 \notag \\
\equiv & \log L_{x|y}(\boldsymbol{\beta} | \mathbf{x}, \mathbf{y}, \boldsymbol{\xi} ),  \label{eq: likelihood_x} \\
\log L_y( b | \mathbf{y}) = & -\frac{T-2}{2} \log \sigma_y^2 - \frac{1}{2 \sigma_y^2} \sum_{t=2}^{T-1} (y_t - b y_{t-1})^2,
\end{align}
where we have omitted the constant terms not related to model parameters, and $\boldsymbol{\xi} = (\xi_{T-1}, \xi_{T-2} , \dots, \xi_2)$.
Though our interest is the likelihood function $L_{x|y}( \boldsymbol{\beta}, b | \mathbf{x}, \mathbf{y} )$, this likelihood is over-parameterized and unidentifiable.
Thus, we cannot determine the model parameters uniquely by using $L_{x|y}( \boldsymbol{\beta}, b | \mathbf{x}, \mathbf{y} )$ alone.
Therefore, we propose a parameter estimation procedure where we first estimate the parameter $b$ by using maximum likelihood estimation for $L_y(b | \mathbf{y})$ and then estimate the remaining parameters in $L_{x|y}(\boldsymbol{\beta}, b | \mathbf{x}, \mathbf{y})$ by substituting the previously estimated parameter $b = \hat{b}$.
In other words, we first estimate the noise term $\xi_t$ by using maximum likelihood estimation for $L_y(b | \mathbf{y})$, then we substitute the maximum likelihood estimates $\xi_t = \hat{\xi}_t$ for the explanatory variables in $L_{x|y}(\boldsymbol{\beta} | \mathbf{x}, \mathbf{y}, \boldsymbol{\xi} )$.
The justification for this parameter estimation procedure is given as follows.
Since we assume the asymmetric network structure, the model parameter in $L_y(b | \mathbf{y})$ can be estimated independently on $x_t$.
That is, the model parameter $b$ can be estimated by $y_t$ alone, and we can obtain the residual time series $\hat{\xi}_t  \equiv y_t - \hat{b} y_{t-1}$.
Inserting the residual time series $\hat{\xi}_t$ into the likelihood function $L_{x|y}(\boldsymbol{\beta} | \mathbf{x} ,\mathbf{y}, \boldsymbol{\xi})$ reduces the model for $x_t$ to a simple AR model with the explanatory variables $\hat{\xi}_t$ added, as shown in Eq.~(\ref{eq: likelihood_x}).
Therefore, the maximum likelihood estimate of the inserted likelihood $L_{x|y}( \boldsymbol{\beta} | \mathbf{x}, \mathbf{y},  \hat{\boldsymbol{\xi} })$ always increases by including interaction term $c$.

Partial Granger causality in the literature suffers from a negative value of the test statistics.
In Ref.~\cite{dpgc}, it was pointed out that the reason for this may be due to small coding errors.
We infer the reason for this from the discussion of the parameter estimation of the model.
In the method of partial Granger causality, maximum likelihood estimation for joint distribution $p(x,z)$ was proposed as a method of parameter estimation~\cite{Barrett_2010}, where $x$ and $z$ are the target variable and the conditioning variable, respectively.
Even in our Granger causality measure, maximum likelihood estimation for the overall likelihood $L$ is a possible candidate for the parameter estimation.
However, when we use maximum likelihood estimation for $L$, $L$ always increases by including the interaction term although $L_{x|y}$ does not necessarily increase.
It is because, the interaction parameter included is used to increase $L$ itself, but not increase $L_{x|y}$.
We have numerically confirmed that only occasionally maximum likelihood estimation for $L$ may lead to a negative value of the test statistics.

If the noise term $\xi_t$ were observed, the distribution of test statistics for inferring directed interactions can be obtained analytically based on the log-likelihood ratio statistic in the framework of the generalized linear model as follows.
We denote the model parameters of our interest as $\boldsymbol{\beta}_1 = (a, c, \eta)$, and the parameters of a full model as $\boldsymbol{\beta}_{\mathrm{max}}$.
The full model is a generalized linear model with the same probability distribution and the same link function as the model of our interest, except that it has a maximum number of parameters that can be estimated from the data.
The number of parameters for the model of our interest and the full model is $p_1$ and $T-1$, respectively.
Since the likelihood function $L_{x|y}(\boldsymbol{\beta} | \mathbf{x}, \mathbf{y}, \boldsymbol{\xi} )$ is expressed as a product of that of the Gaussian distribution, the deviance statistic $D_1$ obeys the chi-square distribution, $D_1 \sim \chi^2(T-1-p_1)$, provided that the model of our interest describes the data as much as the full model~\cite{Dobson,Mccullagh}:
\begin{align}
D_1 \equiv& 2 \bigl[ \log L_{x|y}(\hat{\boldsymbol{\beta}}_{\mathrm{max}} | \mathbf{x}, \mathbf{y}, \boldsymbol{\xi} )  - \log L_{x|y}(\hat{\boldsymbol{\beta}}_1  | \mathbf{x}, \mathbf{y}, \boldsymbol{\xi}  ) \bigr]  \notag \\
=& \frac{1}{\tau^2} \sum_{t=2}^T (x_t - \hat{\mu}_t^{(1)})^2 ,
\end{align}
where
\begin{align}
\hat{\mu}_t^{(1)}
=& \hat{a} x_{t-1} + \hat{c} y_{t-1} + \hat{\eta} \xi_{t-1},
\end{align}
and hats denote maximum likelihood estimates for $L_{x|y}(\boldsymbol{\beta} | \mathbf{x}, \mathbf{y}, \boldsymbol{\xi})$.
On the other hand, if there is a significant difference in describing the data between model of our interest and the full model, the deviance obeys the non-central chi-square distribution and thus, takes a larger value than that expected by the central chi-square distribution.

When we want to test the existence of directed interactions, we set a null hypothesis $H_0$ and an alternative hypothesis $H_1$.
Model parameters corresponding to these hypotheses are expressed as $\boldsymbol{\beta}_0 = (a, \eta)$ and $\boldsymbol{\beta}_1 = (a, c, \eta)$, respectively.
These model are nested, that is, they have the same probability distribution and the same link function, but the parameters of the model for $H_0$ is a special case of the parameters of the model for $H_1$.
The number of parameters for these models is $p_0$ and $p_1$, respectively.
Then, we consider the difference of deviance between these models,
\begin{align}
\Delta D \equiv &  D_0 - D_1 \notag \\
=& 2 \bigl[ \log L_{x|y}(\hat{\boldsymbol{\beta}}_1  | \mathbf{x}, \mathbf{y}, \boldsymbol{\xi}  ) - \log L_{x|y}(\hat{\boldsymbol{\beta}}_0  | \mathbf{x}, \mathbf{y}, \boldsymbol{\xi}  ) \bigr] \notag \\
=& \frac{1}{ \tau^2} \sum_{t=2}^T \Bigl[ (x_t - \hat{\mu}_t^{(0)} )^2 - (x_t - \hat{\mu}_t^{(1)} )^2 \Bigr],
\end{align}
where
\begin{align}
D_0 =& \frac{1}{ \tau^2} \sum_{t=2}^T (x_t - \hat{\mu}_t^{(0)})^2, \\
\hat{\mu}_t^{(0)}
=& \hat{a} x_{t-1} + \hat{\eta} \xi_{t-1},
\end{align}
and hats denote maximum likelihood estimates for $L_{x|y}(\boldsymbol{\beta} | \mathbf{x}, \mathbf{y}, \boldsymbol{\xi})$.
The difference of deviance obeys the central chi-square distribution, $\Delta D \sim \chi^2(p_1 - p_0)$, provided that both models describe the data as much as the full model.
On the other hand, if there is a significant difference in describing the data between these models, the difference of deviance obeys the non-central chi-square distribution, and takes a larger value than that expected by the central chi-square distribution.

Here, we are assuming that the dispersion parameter of the model $\tau$ cannot be estimated from the data.
Then, the difference of deviance is proportional to unknown parameter $\tau$, i.e., the difference of deviance is expressed as a scaled deviance.
In practice, we can remove this unknown parameter in the test statistics by dividing deviances.
First, we assume that the model for the alternative hypothesis describes the data as much as the full model.
That is, the deviance $D_1$ obeys the central chi-square distribution.
Next, we divide the difference of deviance $\Delta D$ by $D_1$ to make a new statistic that is independent of the unknown parameter $\tau$,
\begin{align}
F =& \frac{\Delta D /(p_1 - p_0) }{D_1/(T-1 - p_1)} \notag \\
=& \frac{ \frac{1}{p_1 - p_0} \sum_{t=2}^{T}  \Bigl[ (x_t - \hat{\mu}_t^{(0)} )^2 - (x_t - \hat{\mu}_t^{(1)} )^2 \Bigr] }
{ \frac{1}{T-1-p_1} \sum_{t=2}^T (x_t - \hat{\mu}_t^{(1)})^2 }.
\end{align}
The statistic $F$ obeys the $F$-distribution, $F \sim  F(p_1 - p_0, T-1  - p_1)$, provided that both models describe the data as much as the full model.
On the other hand, if there is a significant difference in describing the data between the null model and the alternative model, the statistic $F$ obeys the non-central $F$-distribution and takes a larger value than that expected by the central $F$-distribution.
In summary, the existence of directed interactions can be tested as follows.
If the test statistic $F$ is in the significance level of the $F$-distribution, we judge that there is no difference in describing the data between these models and accept a simpler model $H_0$.
If the test statistic $F$ takes a value larger than the significance level, we reject $H_0$ and accept $H_1$.

The above result strictly holds true if $\xi_t$ were observed.
In fact, $\xi_t$ is not observed and has to be estimated from the data.
Here, we approximate $\xi_t$ by its maximum likelihood estimates for $L_y(b | \mathbf{y})$:
\begin{align}
L_{x|y}(\boldsymbol{\beta} | \mathbf{x}, \mathbf{y}, \boldsymbol{\xi}) \simeq & \ L_{x|y}(\boldsymbol{\beta} | \mathbf{x}, \mathbf{y}, \hat{\boldsymbol{\xi}}), \\
\hat{\xi}_t \equiv & \  y_t - \hat{b} y_{t-1}.
\end{align}
Then, model parameters $\boldsymbol{\beta}$ are estimated from the approximated likelihood $L_{x|y}(\boldsymbol{\beta} | \mathbf{x}, \mathbf{y}, \hat{\boldsymbol{\xi}})$.
In this paper, we assume that the test statistic obeys the $F$-distribution approximately even if we substitute $\hat{\xi}_t$.

The above result can be directly extended to a more general model where the model order of the interacting term and the lag of the noise correlation take any values.
In this case, the model of the time series is expressed as follows:
\begin{align}
x_t = & \sum_{i=1}^{l_a} a_i x_{t-i} + \sum_{i=1}^{l_c} c_i x_{t-i} + \varepsilon_t \notag \\
=& \sum_{i=1}^{l_a} a_i x_{t-i} + \sum_{i=1}^{l_c} c_i x_{t-i} + \eta_{l_{\eta}} \xi_{t - l_{\eta}} + \omega_t, \\
y_t = & \sum_{i=1}^{l_b} b_i y_{t-i} + \xi_t,
\end{align}
We express the model parameters as $\boldsymbol{\beta}=(\mathbf{a}, \mathbf{c}, \eta_{l_{\eta}})$, where $\mathbf{a}=(a_1, a_2, \cdots, a_{l_a})$, $\mathbf{c}=(c_1, c_2, \cdots, c_{l_c})$ and $l_{\eta}$ denotes the time point of lag for noise correlation and takes an integer value larger than zero.
Then, the log-likelihood functions for this model are given by
\begin{align}
\log L_{x|y}(\boldsymbol{\beta}, \mathbf{b} | \mathbf{x}, \mathbf{y} )
=& -\frac{(T-l_{\mathrm{max}})}{2} \log \tau^2 \notag \\
& - \frac{1}{2 \tau^2} \sum_{t=l_{\mathrm{max} + 1}}^T
\Biggl[ x_t - \sum_{i=1}^{l_a} a_i x_{t-i} - \sum_{i=1}^{l_c} c_i y_{t-i} - \eta_{l_{\eta}} \Bigl( y_{t - l_{\eta}} - \sum_{i=1}^{l_b} b_i y_{t - l_{\eta} -i} \Bigr) \Biggr] \notag \\
=& -\frac{(T-l_{\mathrm{max}})}{2} \log \tau^2 - \frac{1}{2 \tau^2} \sum_{t=l_{\mathrm{max} + 1}}^T
\Biggl[ x_t - \sum_{i=1}^{l_a} a_i x_{t-i} - \sum_{i=1}^{l_c} c_i y_{t-i} - \eta_{l_{\eta}} \xi_{t - l_{\eta}} \Biggr] \notag \\
\equiv & \log L_{x|y}( \boldsymbol{\beta} | \mathbf{x}, \mathbf{y}, \boldsymbol{\xi} ), \\
\log L_y( \mathbf{b} | \mathbf{y} ) 
=& - \frac{(T-1-l_b)}{2} \log \sigma_y^2 - \frac{1}{2 \sigma_y^2} \sum_{t=l_b + 1}^{T-1} \Biggl[ y_t - \sum_{i=1}^{l_b} b_i y_{t-i} \Biggr],
\end{align}
where
\begin{align}
l_{\mathrm{max}} =& \mathrm{max}(l_a, \  l_c, \  l_b+l_{\eta}).
\end{align}

It should be noted that since we are dealing with the AR model, the noise $\varepsilon_t$ and $\xi_t$ does not have autocorrelation.
Therefore, $\varepsilon_t$ and $\xi_t$ cannot correlate with multiple time points.
Thus, we just consider the correlation between $\varepsilon_t$ and $\xi_t$ with a single time point $\eta_{l_{\eta}} \xi_{t-l_{\eta}}$ rather than multiple time points $\sum_{i=1}^{l_{\eta}} \eta_{i} \xi_{t-i}$.
The specific expression for the test statistic $F$ is given by
\begin{align}
F =& \frac{\frac{1}{p_1 - p_0} \sum_{t=l_{\mathrm{max}}+1}^T \Bigl[ (x_t - \hat{\mu}_t^{(0)})^2 - (x_t - \hat{\mu}_t^{(1)})^2 \Bigr] }{ \frac{1}{T-l_{\mathrm{max}}-p_1} \sum_{t=l_{\mathrm{max}} + 1}^T (x_t - \hat{\mu}_t^{(1)} )^2 },
\end{align}
where
\begin{align}
\hat{\mu}_t^{(1)} = & \sum_{i=1}^{l_a} \hat{a}_i x_{t-i} + \sum_{i=1}^{l_c} \hat{c}_i y_{t-i} + \hat{\eta}_{l_{\eta}} \xi_{t-l_{\eta} }, \\
\hat{\mu}_t^{(0)} = & \sum_{i=1}^{l_a} \hat{a}_i x_{t-i} + \hat{\eta}_{l_{\eta}} \xi_{t-l_{\eta} },
\end{align}
and hats denote maximum likelihood estimates for $L_{x|y}(\boldsymbol{\beta} | \mathbf{x}, \mathbf{y}, \boldsymbol{\xi})$.
The statistic $F$ obeys the $F$-distribution, $F \sim  F(p_1 - p_0, T-l_{\mathrm{max}}  - p_1)$, provided that both models describe the data as much as the full model.

Again, the model parameters $\mathbf{b}$ are calculated by maximum likelihood estimation for $L_y(\mathbf{b} | \mathbf{y})$ first, and the estimated values for $\xi_t$ is obtained as
\begin{align}
\hat{\xi}_t \equiv \ y_t - \sum_{i=1}^{l_b} b_i y_{t-i}.
\end{align}
Then, the likelihood function $L_{x|y}(\boldsymbol{\beta} | \mathbf{x}, \mathbf{y} , \boldsymbol{\xi})$ is approximated by inserting $\xi_t = \hat{\xi}_t$:
\begin{align}
L_{x|y}(\boldsymbol{\beta} | \mathbf{x}, \mathbf{y}, \boldsymbol{\xi}) \simeq & \ L_{x|y}(\boldsymbol{\beta} | \mathbf{x}, \mathbf{y}, \hat{\boldsymbol{\xi}}).
\end{align}
The model parameters $\boldsymbol{\beta}$ are estimated from the approximated likelihood $L_{x|y}(\boldsymbol{\beta} | \mathbf{x}, \mathbf{y}, \hat{\boldsymbol{\xi}})$, and we assume that the approximated test statistic approximately obeys the $F$-distribution, even if we substitute $\hat{\xi}_t$.

\section{Numerical experiments\label{sec: numerical_experiment}}
In this section, we confirm the validity of our parameter estimation procedure described in the previous section and the distribution that the test statistic $F$ obeys numerically.
Then, we propose a practical procedure for inferring directed interactions, which is based on a multiple statistical test for the case where the influence of latent common inputs may exist.
In all numerical experiments in this paper, we assume that the model order of all the self interactions $l_a$ and $l_b$ is known for simplicity.
In addition, we assume that the network structure is asymmetric, i.e., the interaction from $x_t$ to $y_t$ is absent.

\begin{figure}[htbp]
   \begin{tabular}{c}
      \begin{minipage}{0.25\hsize}
      \centering
      \includegraphics[width=3.5cm]{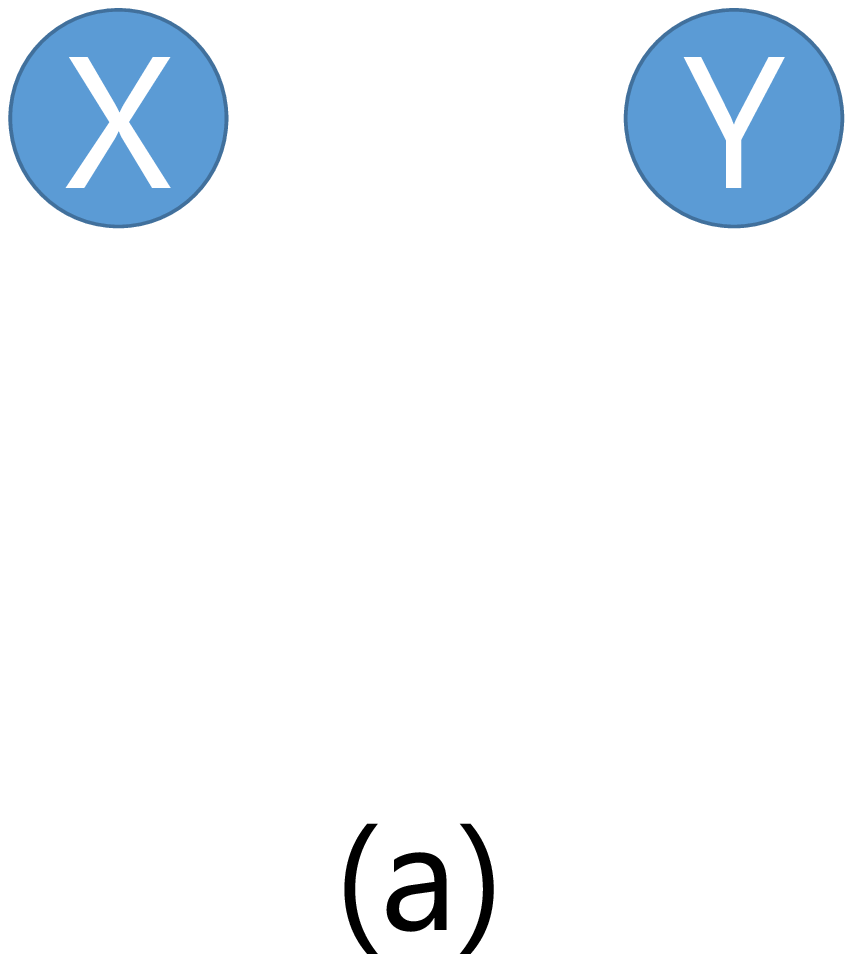}
      \end{minipage}
\begin{minipage}{0.25\hsize}
      \centering
      \includegraphics[width=3.5cm]{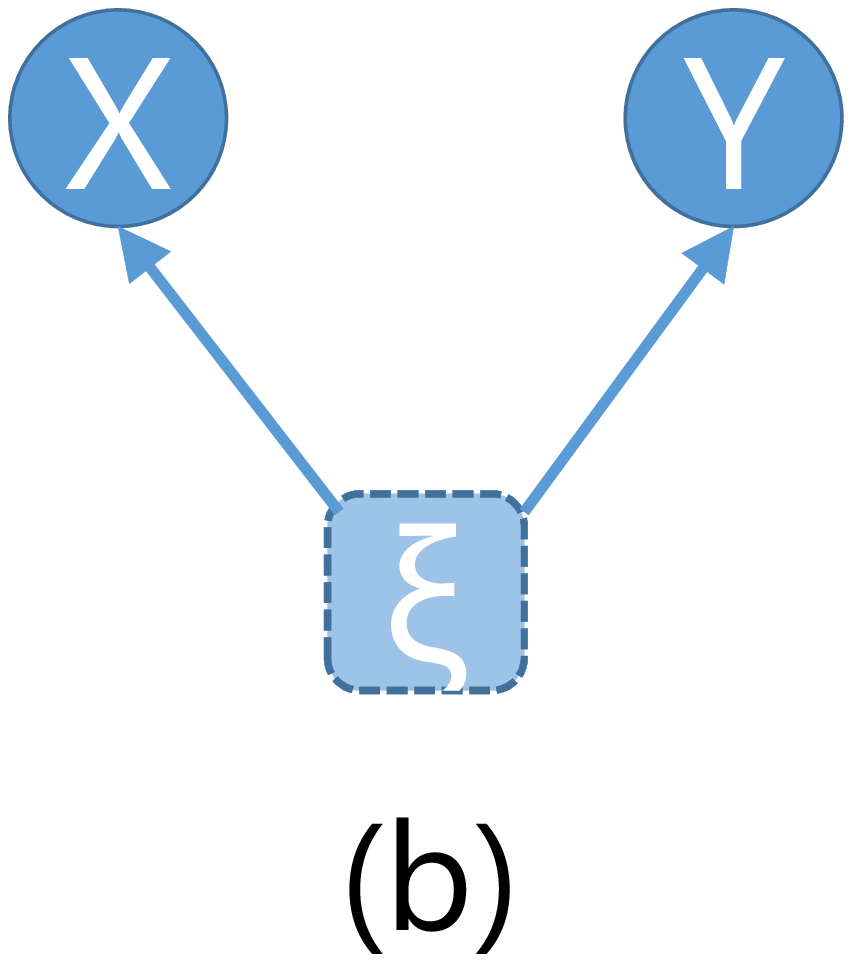}
      \end{minipage}
\begin{minipage}{0.25\hsize}
      \centering
      \includegraphics[width=3.5cm]{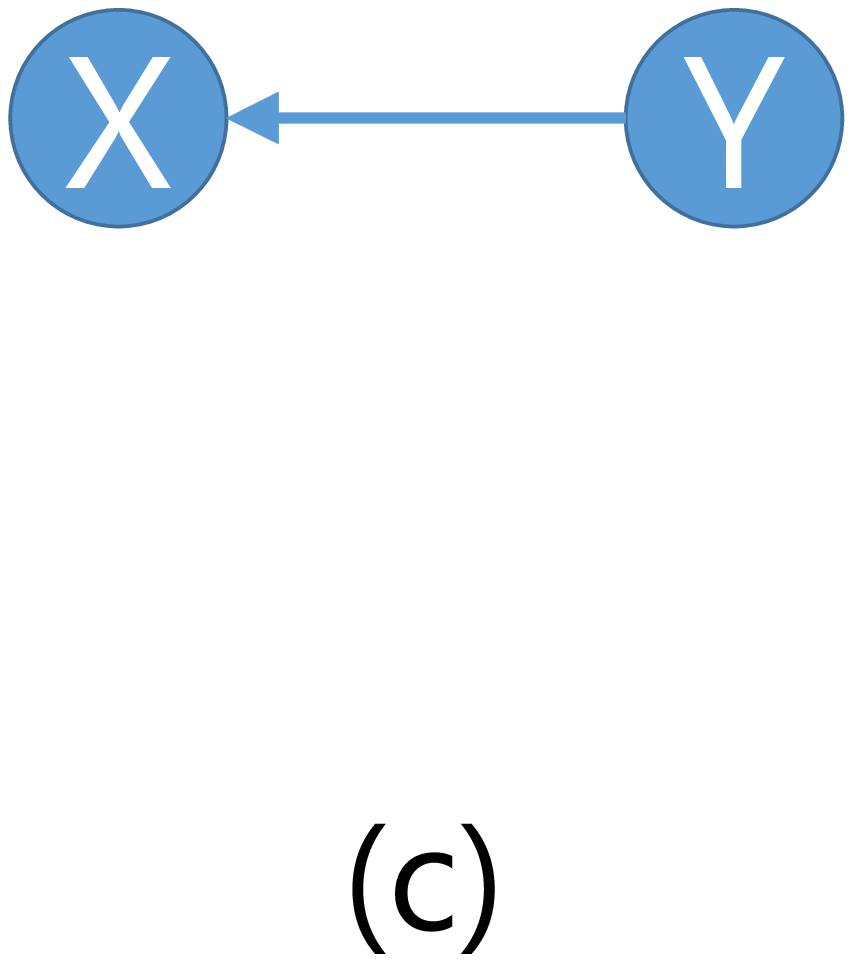}
      \end{minipage}
      \begin{minipage}{0.25\hsize}
      \centering
      \includegraphics[width=3.5cm]{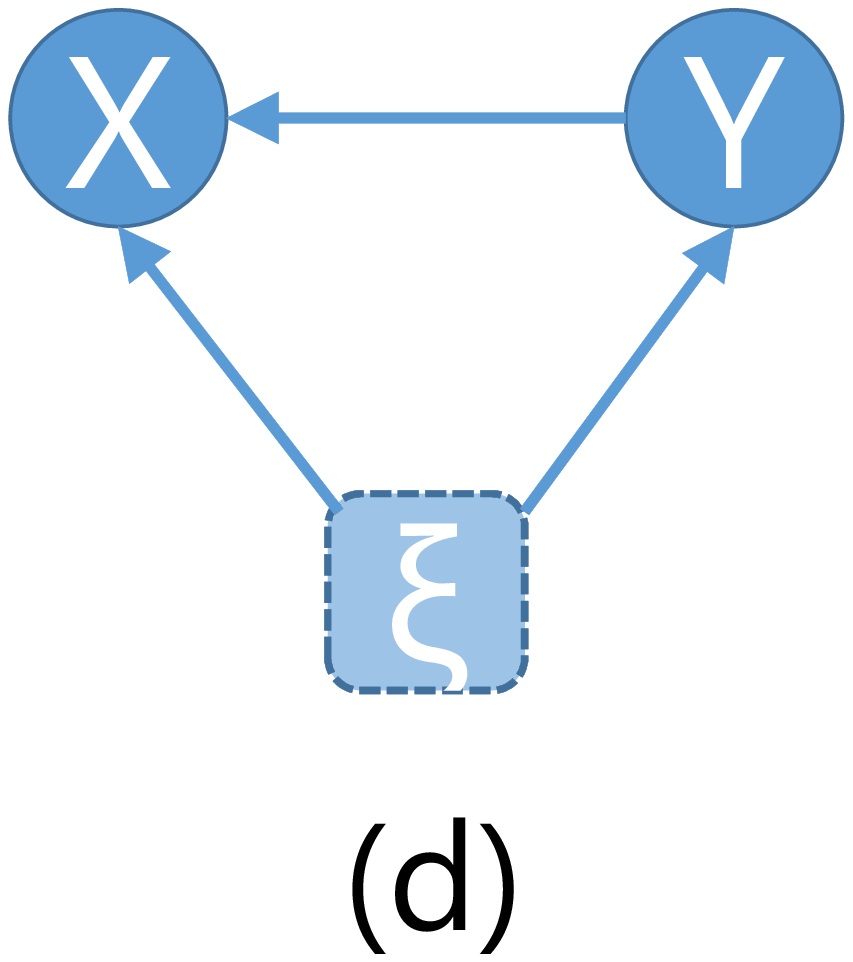}
      \end{minipage}
   \end{tabular}
   \caption{Schematic view of simulated models.}
   \label{fig: simulation_models}
\end{figure}

\begin{figure}[htbp]
   \centering
   \includegraphics[width=13cm]{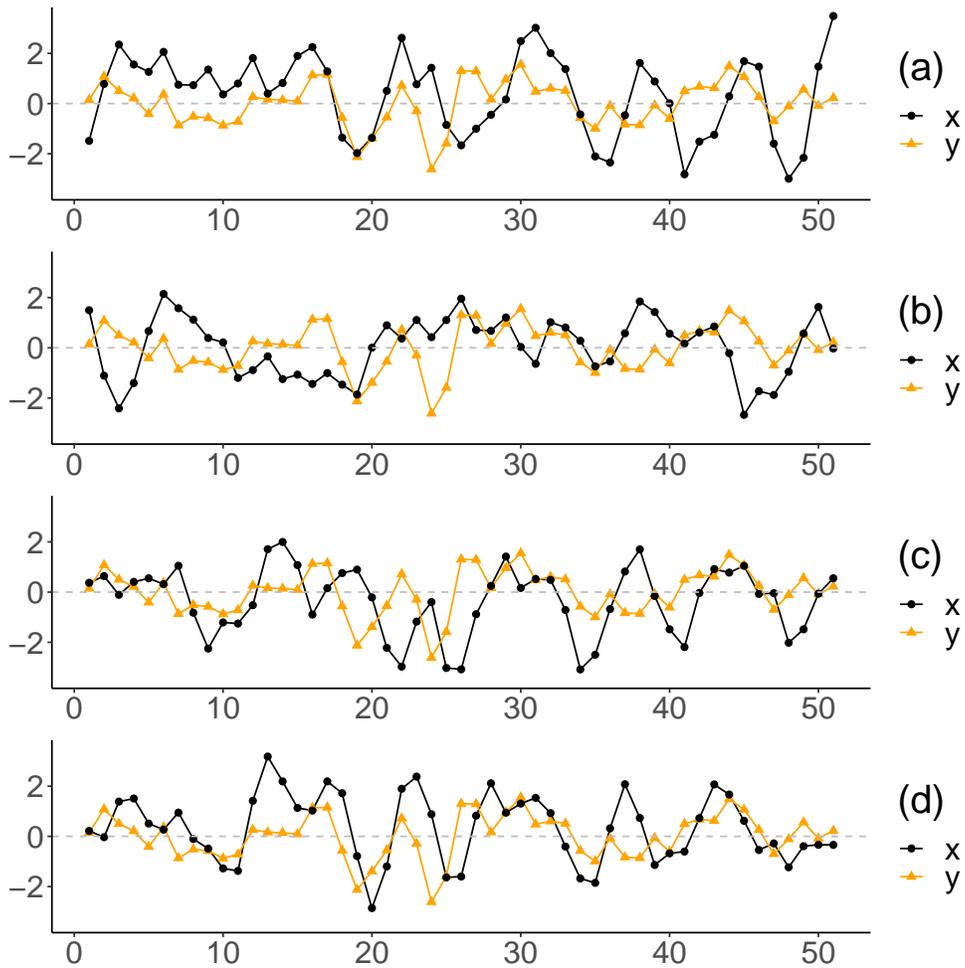}
   \caption{Run-sequence plot for each simulated model. Black lines with circle points denote $x_t$ and orange lines with triangle points denote $y_t$.}
   \label{fig: run_sequence}
\end{figure}

We consider toy models which have been extensively applied in tests of Granger causality~\cite{Baccal_2001,appl_book}, except that we modify this model by adding latent common inputs to each time series.
We simulated four toy models with and without latent common inputs and directed interactions, $(a)$, $(b)$, $(c)$ and $(d)$:
\begin{align}
x_t =& a_1 x_{t-1} + a_2 x_{t-2} + c_1 y_{t-1} + c_2 y_{t-2} + \varepsilon_t, \\
y_t =& b_1 y_{t-1} + b_2 y_{t-2} + \xi_t,
\end{align}
where the noise has a following covariance matrix:
\begin{align}
\Sigma \equiv&  \begin{pmatrix} \mathrm{var}(\varepsilon_t) & \mathrm{cov}(\varepsilon_t, \xi_{t-1}) \\ \mathrm{cov}(\xi_{t-1}, \varepsilon_t) & \mathrm{var}(\xi_{t-1}) \end{pmatrix} \notag \\
= & \begin{pmatrix} \sigma_x^2 & \rho \sigma_x \sigma_y \\ \rho \sigma_x \sigma_y & \sigma_y^2 \end{pmatrix}.
\end{align}
The parameters $a_1$, $a_2$, $b_1$, $b_2$, $\sigma_x$ and $\sigma_y$ take the same value at each model, however, the other parameters take different values.
The values of the model parameters are listed in TABLE~\ref{table: parameters}.
The schematic view of the simulated models and the run-sequence plots of time series are shown in FIG.~\ref{fig: simulation_models} and FIG.~\ref{fig: run_sequence}, respectively.

\begin{table}[htb]
\caption{The values of the model parameters for each simulated model}
\begin{tabular}{ | >{\centering}p{1.5cm} | r | r | r | r | r | r | r | r | r |  } \hline
model         & 
\hspace{0.2cm}  $a_1$ \hspace{0.2cm} &
\hspace{0.2cm}  $a_2$ \hspace{0.2cm}  &
\hspace{0.25cm}   $b_1$ \hspace{0.25cm}  &
\hspace{0.25cm}   $b_2$ \hspace{0.25cm}  & 
\hspace{0.2cm}  $c_1$ \hspace{0.2cm}   & 
\hspace{0.2cm}  $c_2$ \hspace{0.2cm}  &
\hspace{0.2cm}   $\sigma_x$ \hspace{0.2cm}  &
\hspace{0.2cm}   $ \sigma_y$ \hspace{0.2cm}  &
\hspace{0.26cm}   $\rho$ \hspace{0.26cm}   \\ 
\hline \hline 
$(a)$   & 0.9 & $-0.5$ & 0.5  & $-0.2$ &  0  & 0  &  1 & $\sqrt{0.7}$  & 0   \\ \hline
$(b)$  & 0.9 & $-0.5$ & 0.5 & $-0.2$ &  0 & 0 &  1 & $\sqrt{0.7}$ &  0.4   \\ \hline
$(c)$  & 0.9 & $-0.5$ & 0.5 & $-0.2$ &  0.16 & $-0.2$ &1 & $\sqrt{0.7}$ & 0   \\ \hline
$(d)$  & 0.9 & $-0.5$ & 0.5 & $-0.2$ &  0.16 & $-0.2$ &  1 & $\sqrt{0.7}$ &  0.4  \\ \hline
\end{tabular}
\label{table: parameters}
\end{table}

\subsection{Sampling distribution of the test statistics}
\begin{figure}[htbp]
   \begin{tabular}{c}
      \begin{minipage}{0.5\hsize}
   \centering
   \includegraphics[width=7cm]{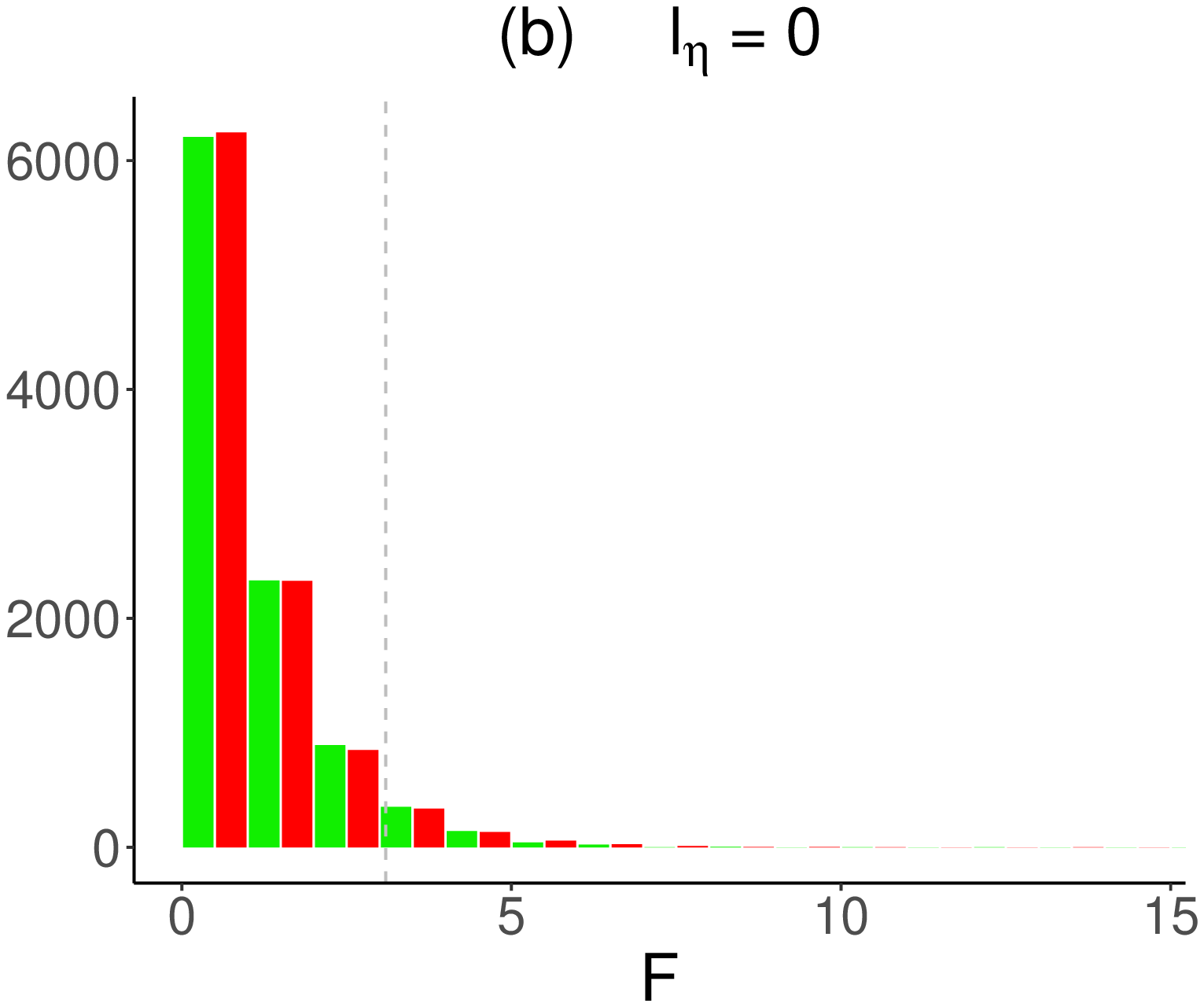}
      \end{minipage}
      \begin{minipage}{0.5\hsize}
   \centering
   \includegraphics[width=7cm]{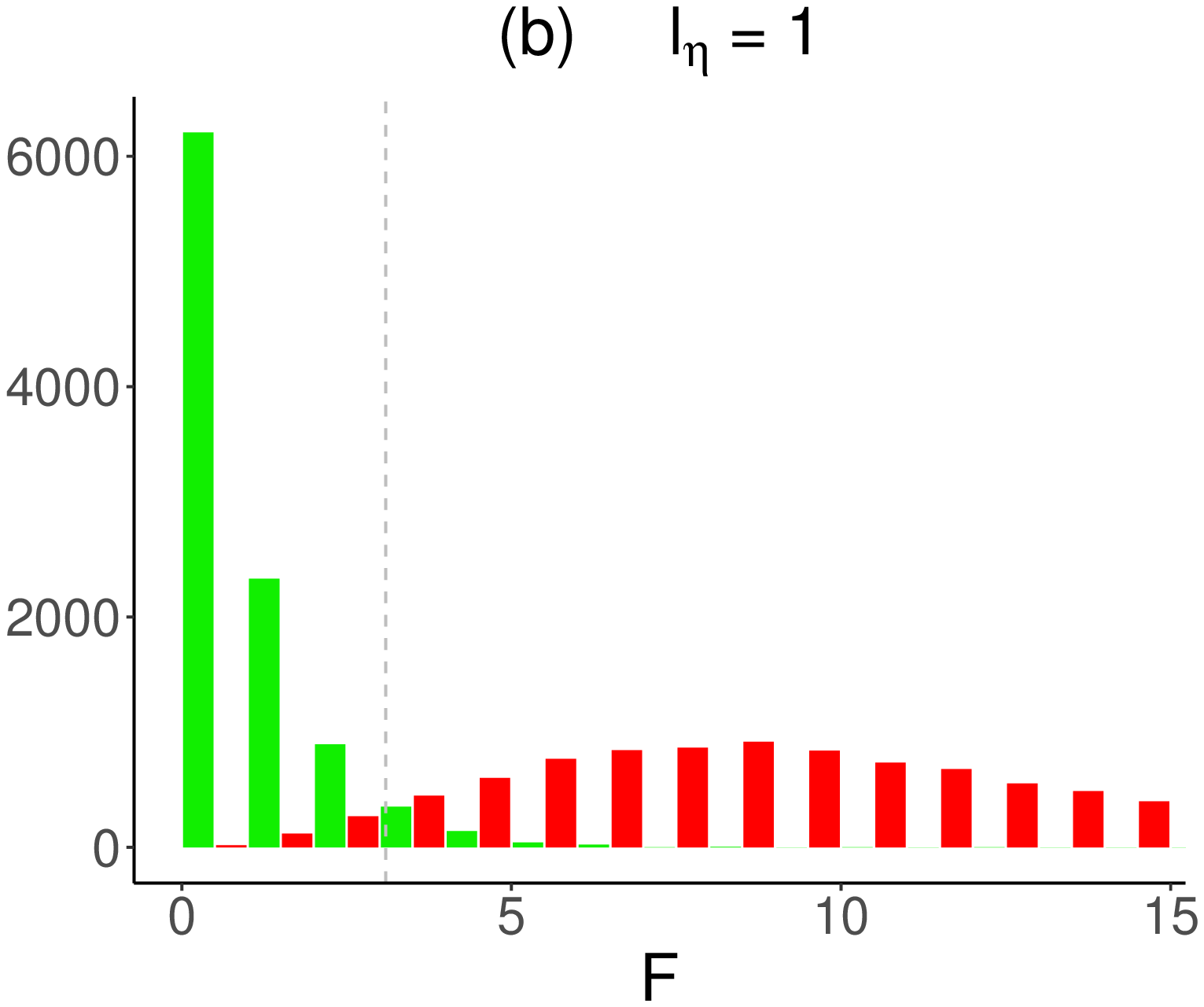}
      \end{minipage}
   \end{tabular}
   \caption{The distributions of test statistics for normal Granger causality under the influence of correlated noise (the red histograms). Time series are simulated using the model $(b)$ but modified so that the noise may have correlations at various time lags. Equal-time correlated noise (left) and different-time correlated noise (right). These are the results for a sample size of 100 with 10,000 trials. The light green histograms are sampled distributions from the $F$-distribution for reference, and the dashed lines denote the 95 percent quantile of the $F$-distribution.}
   \label{fig: test_stat_normal_proposed}
\end{figure}

In this subsection, we first confirm the influence of the correlated noise generated by latent common inputs numerically.
For simplicity, we assume that the model order of directed interactions and the lag of correlated noise are known.
For the case where these lags are unknown will be treated in the next subsection.
We first confirm that the latent common inputs with delays produce a spurious directed interaction while the equal-time correlated noise does not, though only the latent common inputs without delays were considered in the literature.
FIG.~\ref{fig: test_stat_normal_proposed} shows the effect of delays of latent common inputs on the sampling distributions of the test statistics for normal Granger causality.
The Figures show that when latent common inputs reach simultaneously, the test statistic is the same as that of the null hypothesis $F$-distribution, while in the case of latent common inputs with delay the test statistic deviates from the null distribution.
The purpose of this paper is to construct a robust statistical measure for Granger causality against the latent common inputs with delays.

\begin{figure}[htbp]
   \begin{tabular}{c}
      \begin{minipage}{0.333\hsize}
   \centering
   \includegraphics[width=5cm]{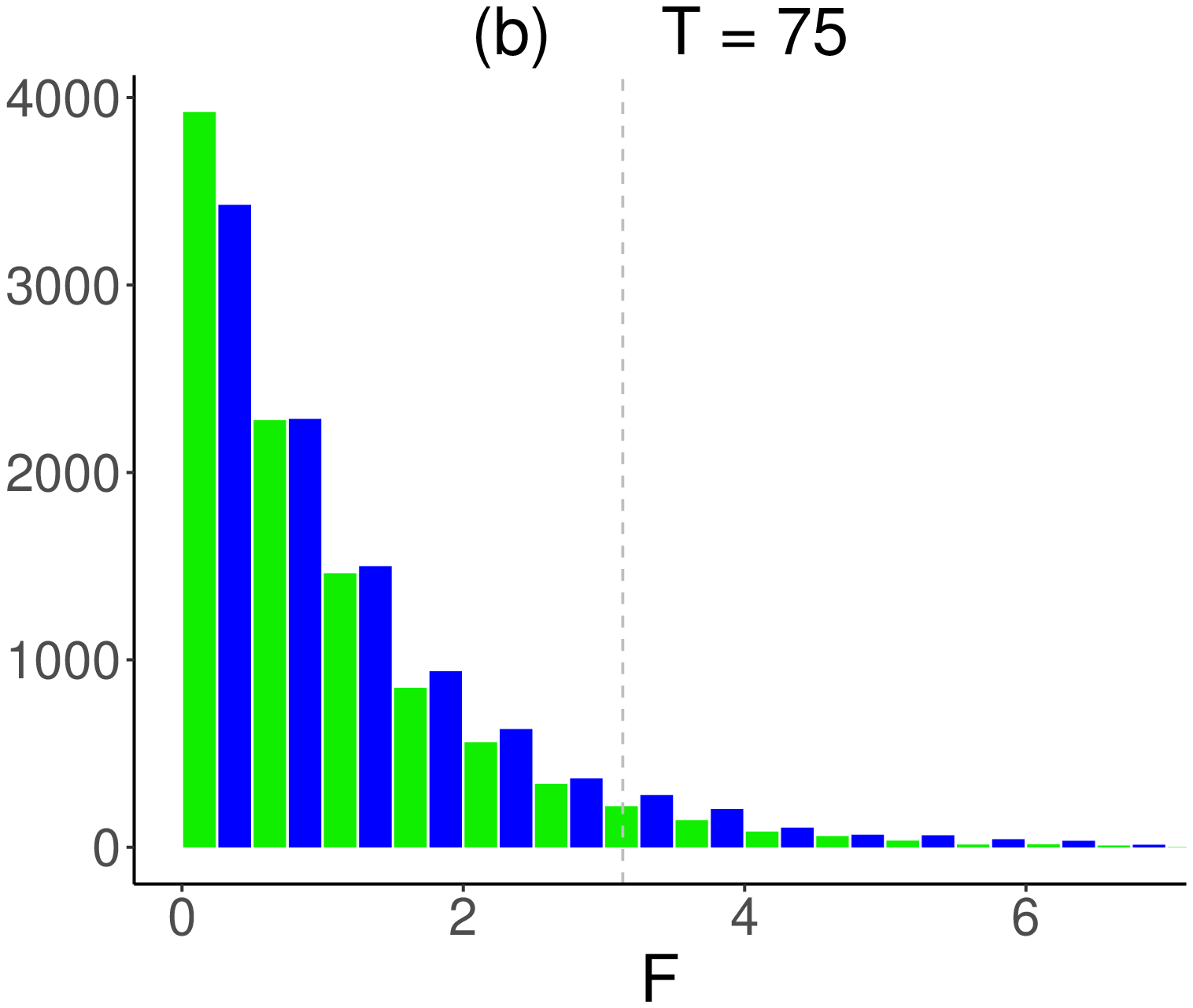}
      \end{minipage}
      \begin{minipage}{0.333\hsize}
   \centering
   \includegraphics[width=5cm]{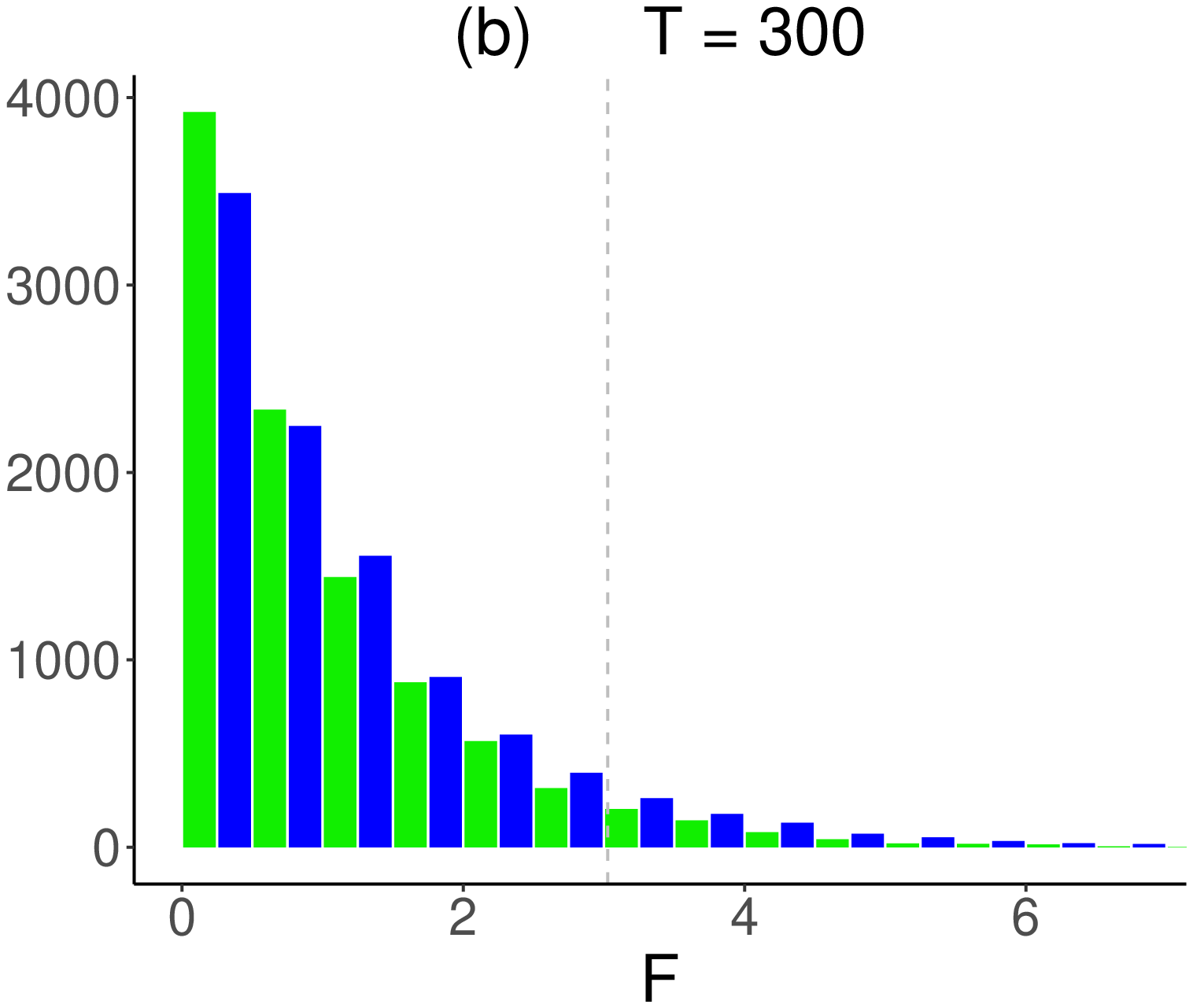}
      \end{minipage}
\begin{minipage}{0.333\hsize}
   \centering
   \includegraphics[width=5cm]{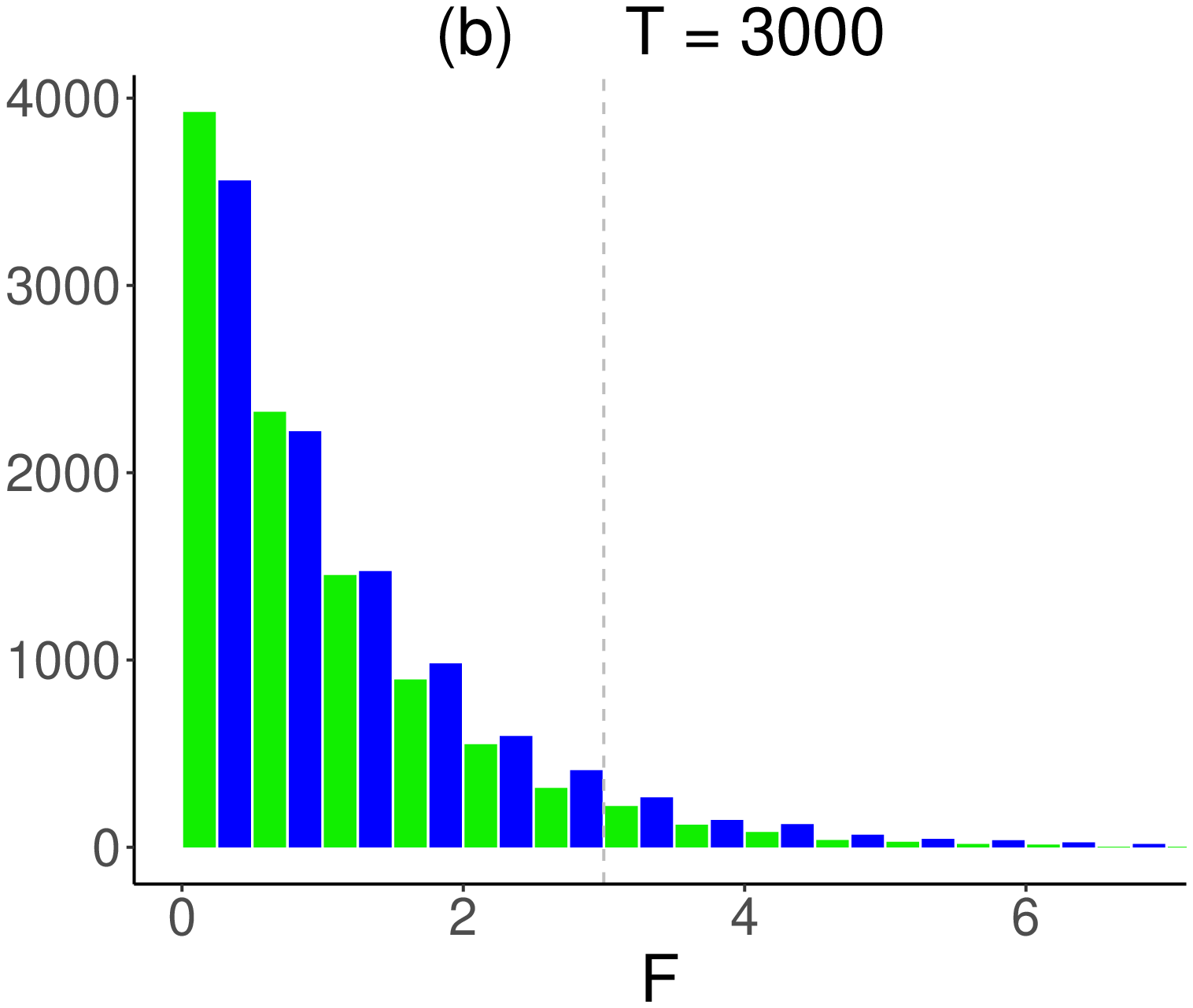}
      \end{minipage}
   \end{tabular}
   \caption{The distributions of test statistics for proposed method (the blue histograms). Results of simulation model $(b)$ with 10,000 trials. Sample sizes are 300 (left), 1,000 (center) and 3,000 (right). The light green histograms are sampled distributions from the $F$-distribution for reference, and the dashed lines denote 95 percent quantile of the $F$-distribution.}
   \label{fig: test_stat_b_sample_size}
\end{figure}

Next, we confirm the validity of our statistical measure and the parameter estimation procedure proposed in the previous section.
FIG.~\ref{fig: test_stat_b_sample_size} shows the test statistics for proposed Granger causality at model $(b)$ with various sample sizes.
From the figure, the proposed test statistics approximately obey the $F$-distribution, and this property does not change even if the sample size varies.
Thus, the proposed method is robust against the influence of latent common inputs.
The reason for why the test statistic does not exactly obey the $F$-distribution is that we have used the approximated likelihood $L_{x|y}(\boldsymbol{\beta} | \mathbf{x}, \mathbf{y}, \hat{\boldsymbol{\xi}})$ rather than $L_{x|y}(\boldsymbol{\beta} | \mathbf{x}, \mathbf{y}, \boldsymbol{\xi} )$.

\begin{figure}[htbp]
   \begin{tabular}{c}
      \begin{minipage}{0.333\hsize}
   \centering
   \includegraphics[width=5cm]{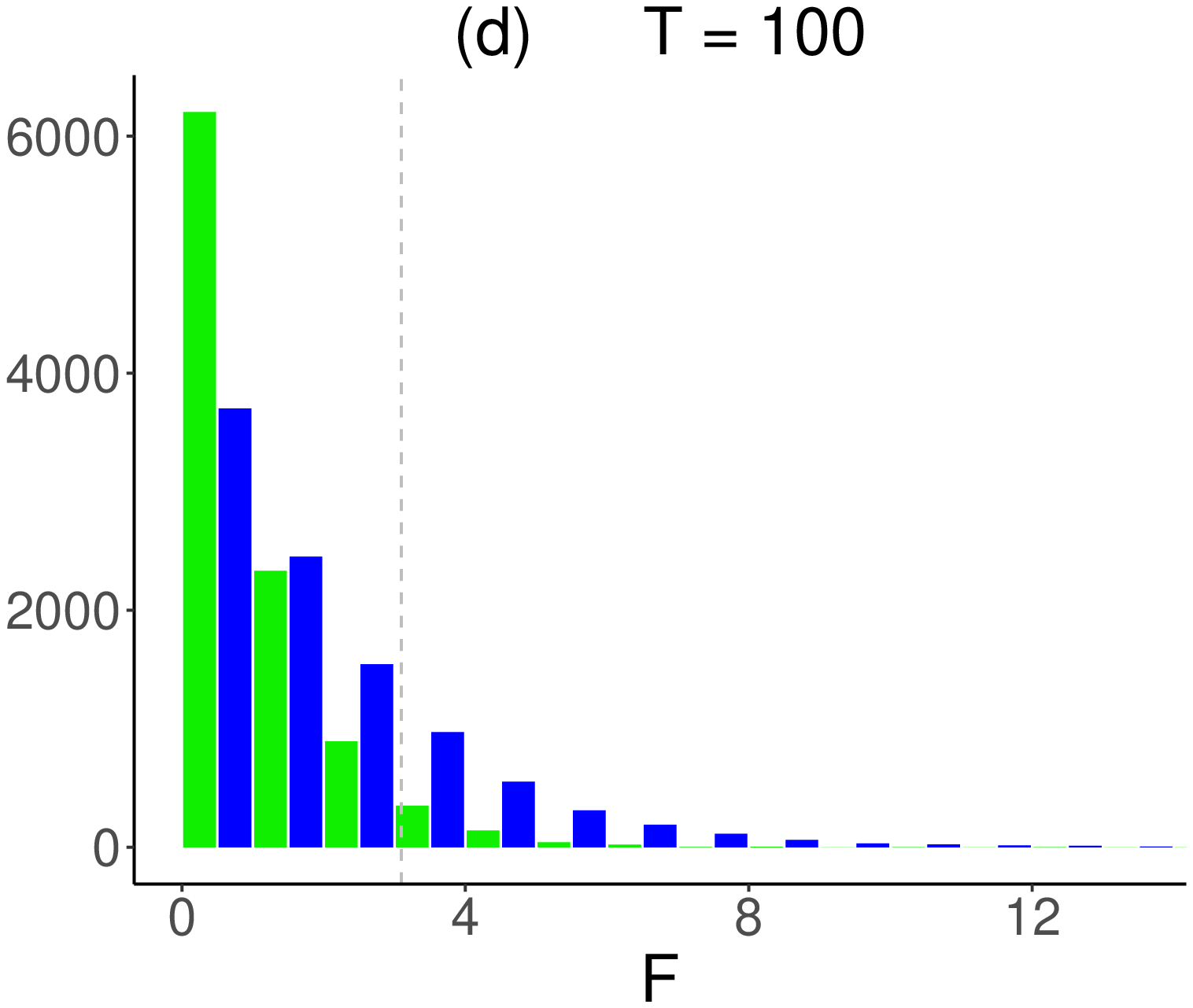}
      \end{minipage}
      \begin{minipage}{0.333\hsize}
   \centering
   \includegraphics[width=5cm]{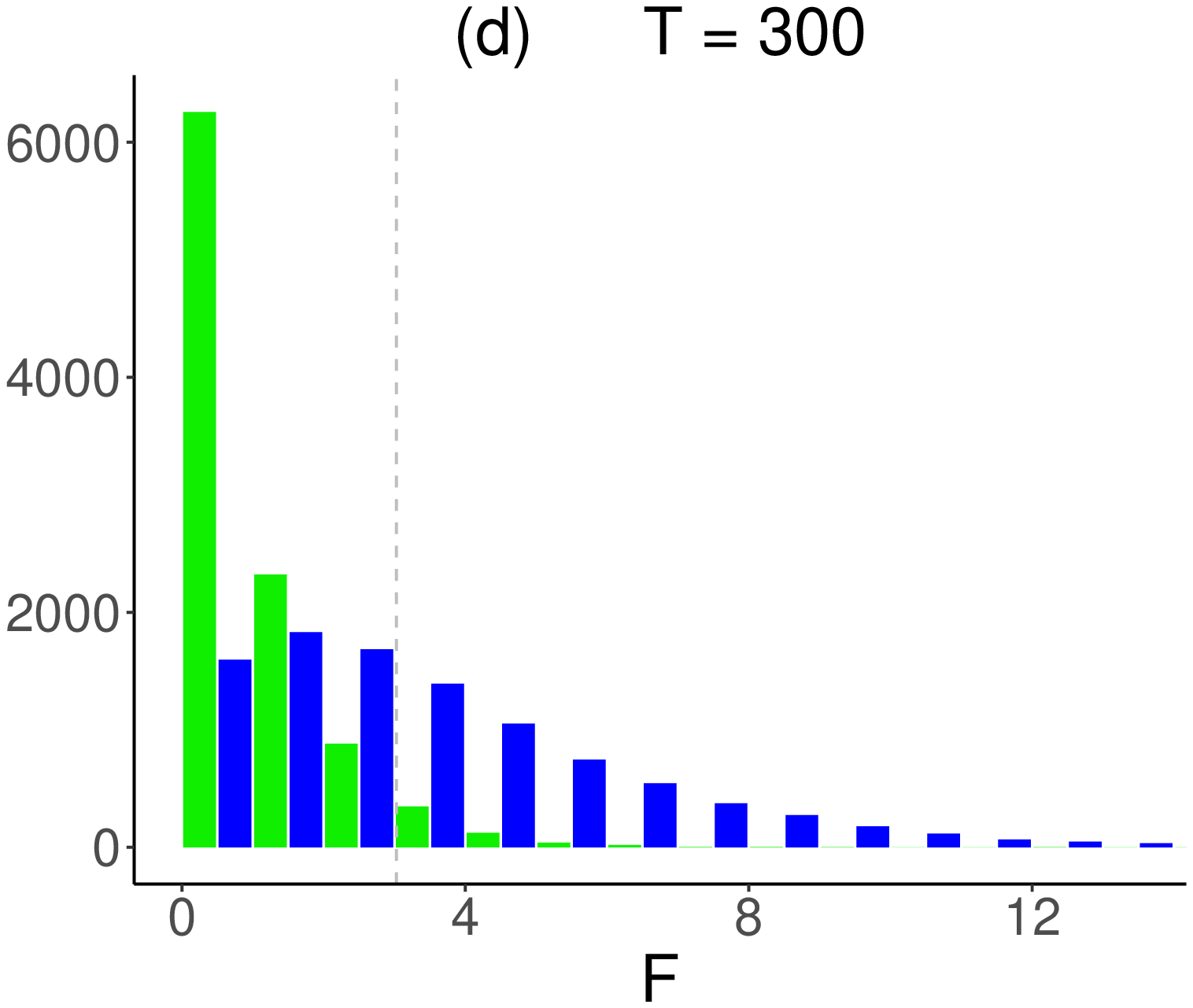}
      \end{minipage}
\begin{minipage}{0.333\hsize}
   \centering
   \includegraphics[width=5cm]{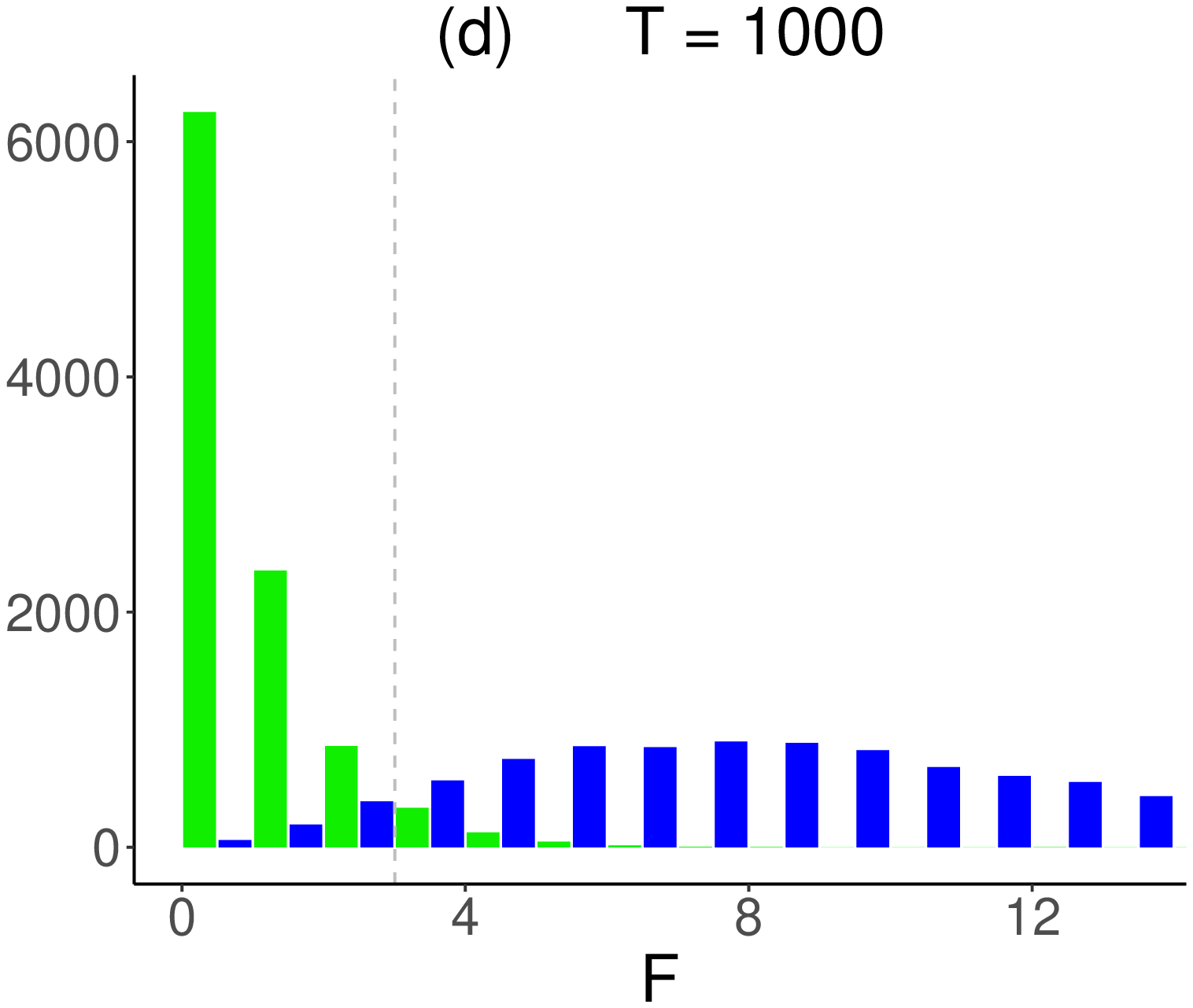}
      \end{minipage}
   \end{tabular}
   \caption{The distributions of test statistics for proposed method (the blue histograms). Result of simulation model $(d)$ with 10,000 trials. Sample sizes are 100 (left), 300 (center) and 1,000 (right). The light green histograms are sampled distributions from the $F$-distribution for reference, and the dashed lines denote 95 percent quantile of the $F$-distribution.}
   \label{fig: test_stat_d_sample_size}
\end{figure}

The above result alone cannot deny the criticism that the proposed method is just insensitive to directed interactions.
Therefore, we show that the proposed method has an ability to detect true directed interactions.
FIG.~\ref{fig: test_stat_d_sample_size} shows the test statistics of simulated model $(d)$ for various sample sizes.
As the sample size increases, the test statistic deviates from the $F$-distribution, thus the method has an ability to detect directed interactions.

\subsection{Practical procedure}
The assumption made in the previous subsection that the model order of auto-regressive coefficients and the lag of correlated noise are known, is not realistic.
In practice, these lags in true models are unknown in many cases.
Furthermore, since the result of a statistical test depends crucially on the number of parameters to be tested, it is necessary to estimate these lags properly in order to infer the interactions accurately. 
Therefore, in this subsection, we propose a practical procedure for testing directed interactions, which is based on a stepwise variable increasing method through multiple tests.

We first estimate model parameter $y_t$ by maximum likelihood estimation for $L_y$, and obtain the residual time series $\hat{\xi}_t$.
Then, we select the best lags of the noise correlation and the directed interaction, $l_{\eta}$ and $l_c$, by the Bayesian Information criterion (BIC) separately.
BIC is given by
\begin{align}
\mathrm{BIC} =& -2 \log L_{x|y} + p \log (T-l) \notag \\
=& (T-l) \log \biggl[ \frac{1}{T-l}\sum_{t=l+1}^T (x_t - \hat{\mu}_t)^2 \biggr] + p \log (T-l),
\end{align}
where $l$ is the maximum number of lags to be searched for, $p$ is the number of model parameters and we have omitted the irrelevant constants which do not depend on the model parameters.
Here we have assumed that when we calculate BIC, the dispersion parameter can be estimated from the data, $\tau = \hat{\tau} = \frac{1}{T-l} \sum_{t=l+1}^T (x_t - \hat{\mu}_t)^2$.
Next, using the selected lags which minimize BIC, the $p$-values for the noise correlation and the directed interaction are calculated separately.
Then, if the $p$-value, which is the smaller one, takes a smaller value than the multiple test threshold, we reject the null hypothesis and adopt the alternative hypothesis.
If either alternative hypothesis is adopted, we again select the lag and perform a statistical test for the other variable under the adopted alternative hypothesis.
Since this procedure can be seen as multiple tests, we apply the Bonferroni correction to the threshold of the $p$-value in order for the familywise error rate to be less than the prescribed threshold.

\begin{figure}[htbp]
   \begin{tabular}{c}
      \begin{minipage}{0.5\hsize}
   \centering
   \includegraphics[width=8cm]{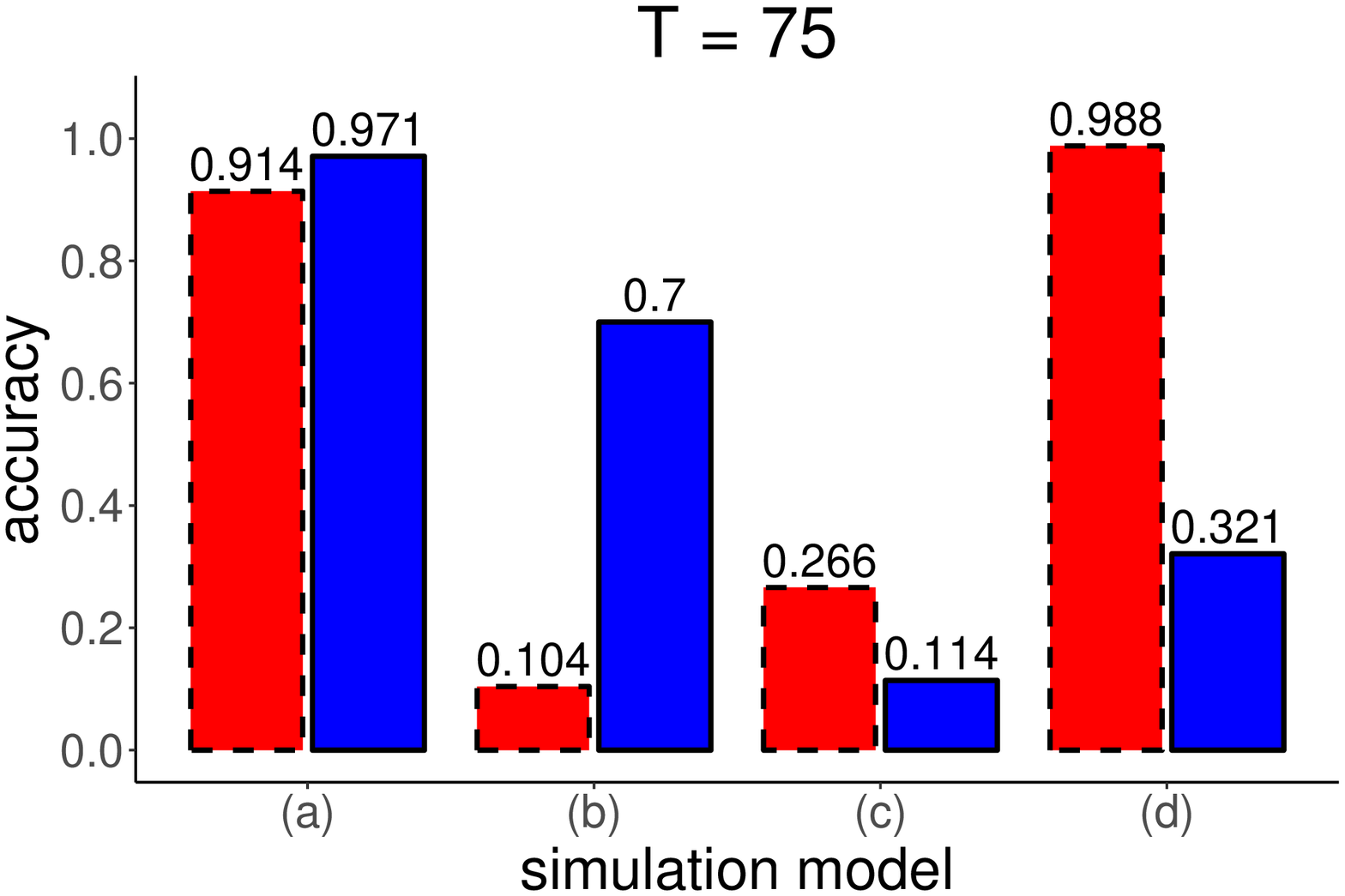}
      \end{minipage}
      \begin{minipage}{0.5\hsize}
   \centering
   \includegraphics[width=8cm]{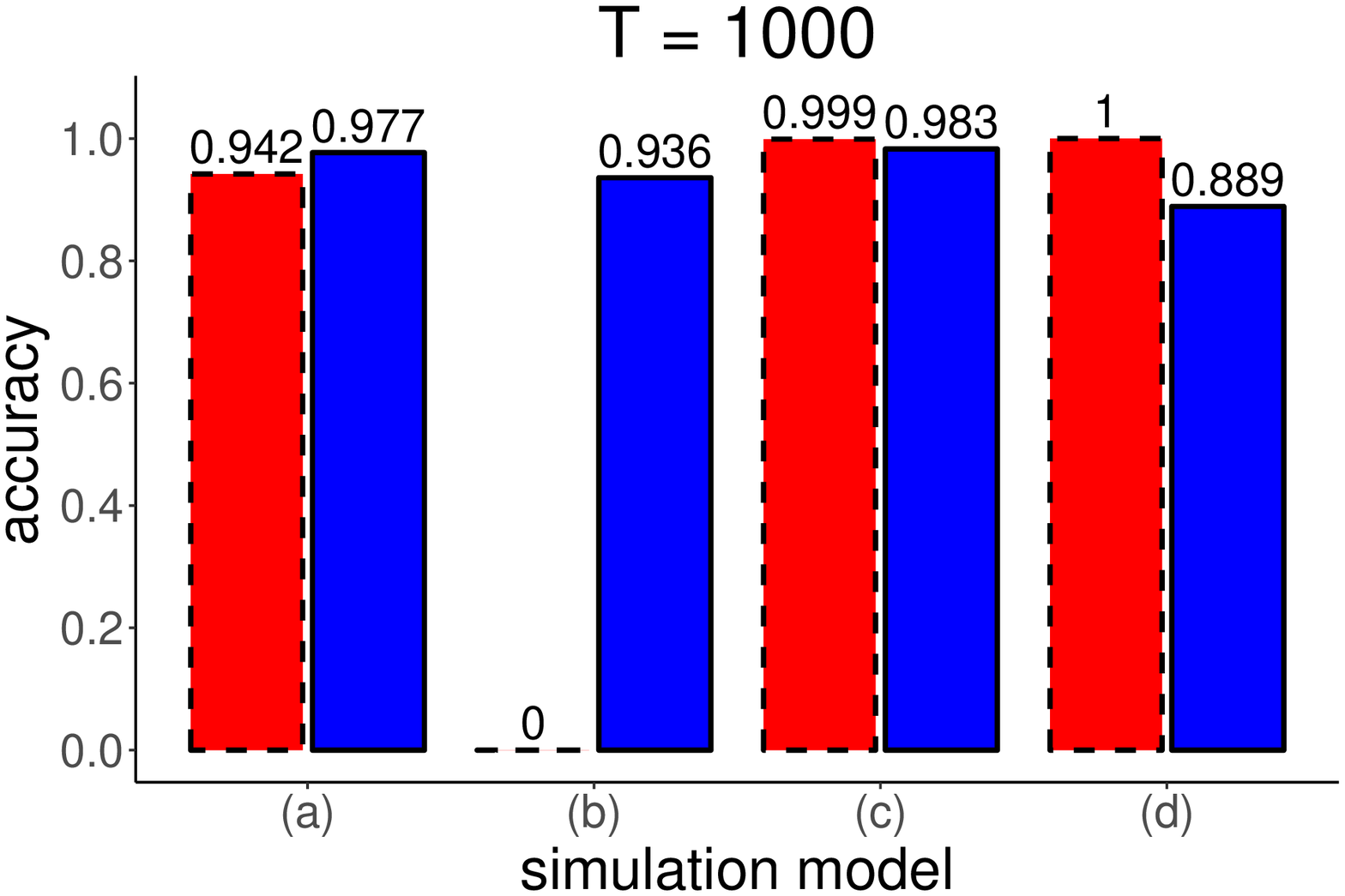}
      \end{minipage}
   \end{tabular}
   \caption{Comparison of accuracy in detecting directed interaction between normal method (red bars with dashed border lines) and proposed method (blue bars with solid border lines). The accuracy is calculated with 10,000 trials for a sample size $T=75$ (left) and $T=1,000$ (right). }
   \label{fig: accuracy}
\end{figure}

We compare the performance in detecting directed interactions between the normal Granger causality procedure and the proposed procedure in FIG.~\ref{fig: accuracy}.
In this numerical experiment and below, we set the maximum number of lags to be searched, $l_{\eta}$ and $l_c$, be 6.
Since the driving force of the AR model is the noise, the correlation between the variable $y_t$ and the noise $\xi_t$, in many cases, is large.
Then, when we add both $y_t$ and $\xi_t$ on explanatory variables when exploring the best lags, the parameter estimation occasionally cannot perform due to multicollinearity.
Therefore, if the multicollinearity occurs, we stop searching for lags.
As in the case of the proposed procedure, we select the best number of lags for the directed interaction and perform a statistical test on it for the normal Granger causality procedure.
We set the threshold of the $p$-value be 0.05 and the Bonferroni correction coefficient to $1/2$ in the case of the multiple test in order the familywise error rate to be 0.05.

From FIG.~\ref{fig: accuracy}, it turns out that when the sample size is small, latent common inputs may produce a spurious directed interaction, even in the proposed method.
It is because, since the noise is the driving force of the AR model, then both $p$-values for interaction $y_t$ and $\xi_t$ take small values, and a comparison of $p$-values between $y_t$ and $\xi_t$ fails.
In addition, the proposed method has a lesser statistical power in detecting the true directed interaction than the normal Granger causality procedure.
However, these shortcomings are resolved by increasing the sample size, and the directed interaction is detectable with high accuracy in any simulated models.
On the other hand, the normal Granger causality procedure detects a spurious directed interaction with a high probability even in the small sample size, and inevitably detects a spurious directed interaction when the sample size is large.

\begin{figure}[htb]
   \centering
   \includegraphics[width=9cm]{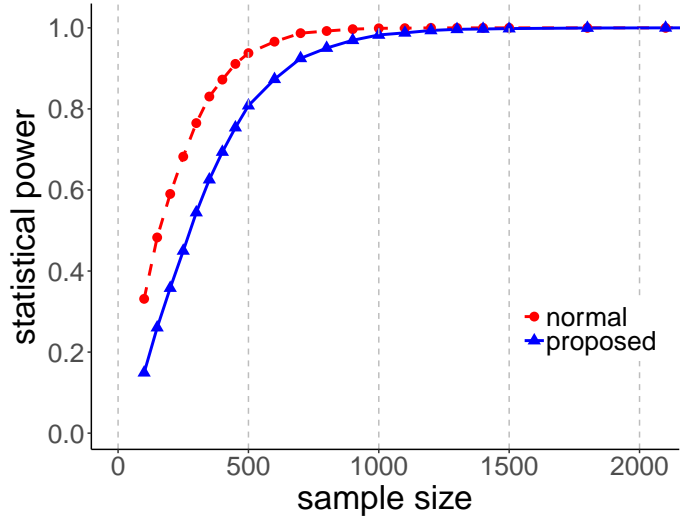}
   \caption{Comparison of statistical power in detecting directed interaction against sample size between normal method (red dashed line with circle points) and proposed method (blue solid line with triangle points). This is a result of simulation model $(c)$ with 10,000 trials.}
   \label{fig: statistical_power}
\end{figure}

In order to quantify the shortcoming of our proposed procedure in detecting the true directed interaction, we compare the statistical power in detecting directed interactions of the simulation model $(c)$ with the normal Granger causality procedure.
FIG.~\ref{fig: statistical_power} plots the statistical power against the sample size for both procedures.
The proposed procedure has always poor statistical power than the normal Granger causality, i.e., the detection performance is conservative.
However, the shortcoming is resolved by increasing the sample size, since the statistical power saturates to 1 as the sample size increases.

\section{Conclusions\label{sec: conclusion}}
The original Granger causality test detects a spurious directed interaction when there are the confounding influence of latent common inputs and latent variables.
In this paper, we propose a new Granger causality measure which is robust against the influence of latent common inputs.
Our method is inspired by partial Granger causality in the literature and its variant, in which the influence of the latent common inputs between the source variable and the target variables was eliminated from the test statistics.
We propose a statistical measure from which the correlated component of the noise between the source variable and the target variable was removed, which was caused by the influence of latent common inputs, while in partial Granger causality the correlated component of the noise between the target variable and the conditioning variable was removed. 
Then, we discuss the sampling distribution of our test statistics by using the likelihood function of time series and show numerically that the statistics approximately obey the $F$-distribution.
According to what the authors in the literature say, the method of eliminating the correlated component of the noise between the target variable and the conditioning variable was called a partial Granger causality, since it was inspired by the partial correlation in statistics.
However, in the light of the discussion in this paper, our method which eliminates the correlated component of the noise between the source variable and the target variable should be said to be a natural extension of Granger causality to the case where the noise has a correlation at a various time lag.

In practice, the model order of interactions and the lag of the correlated noise are unknown.
Therefore, we propose a practical procedure in detecting directed interactions for the case where the influence of latent common inputs may exist, which is based on a stepwise variable increasing method by multiple tests.
The performance of our procedure is verified by using numerically simulated time series.
The normal Granger causality test inevitably detects a spurious directed interaction as the sample size increases, while the proposed method does not.
The robustness of the proposed method against the influence of latent common inputs does not change as the sample size increases, thus our method can remove the influence of the latent common inputs properly.
As a shortcoming of this robustness, the ability of proposed method in detecting the true directed interaction is weaker than the normal Granger causality procedure.
However, this shortcoming can be resolved adequately by increasing the sample size, therefore the proposed method is practical enough.

Of the confounding environmental influence, this study only treats the latent common inputs while the influence of latent variables was not considered.
It is because, the influence of latent variables appears as an auto-correlated noise, which cannot be dealt with in the framework of the AR model strictly.
Granger causality in a model with an auto-correlated noise has to be defined in the framework of the MA model.
The study in this direction has been worked on in the framework of the state space model~\cite{ssgc}.

Our method assumes that the network structure of the interaction between each variable is asymmetric.
Therefore, the extension to the case where the networks structure may be symmetric will be in future works.
In addition, it is important to show a theoretical justification of the parameter estimation proposed in this paper.
Furthermore, it is interesting to extend the proposed method to point process models, such as spike activity model of neuron.

\section*{References}

\bibliography{pgc.bib}

\begin{thebibliography}{24}%
\makeatletter
\providecommand \@ifxundefined [1]{%
 \@ifx{#1\undefined}
}%
\providecommand \@ifnum [1]{%
 \ifnum #1\expandafter \@firstoftwo
 \else \expandafter \@secondoftwo
 \fi
}%
\providecommand \@ifx [1]{%
 \ifx #1\expandafter \@firstoftwo
 \else \expandafter \@secondoftwo
 \fi
}%
\providecommand \natexlab [1]{#1}%
\providecommand \enquote  [1]{``#1''}%
\providecommand \bibnamefont  [1]{#1}%
\providecommand \bibfnamefont [1]{#1}%
\providecommand \citenamefont [1]{#1}%
\providecommand \href@noop [0]{\@secondoftwo}%
\providecommand \href [0]{\begingroup \@sanitize@url \@href}%
\providecommand \@href[1]{\@@startlink{#1}\@@href}%
\providecommand \@@href[1]{\endgroup#1\@@endlink}%
\providecommand \@sanitize@url [0]{\catcode `\\12\catcode `\$12\catcode
  `\&12\catcode `\#12\catcode `\^12\catcode `\_12\catcode `\%12\relax}%
\providecommand \@@startlink[1]{}%
\providecommand \@@endlink[0]{}%
\providecommand \url  [0]{\begingroup\@sanitize@url \@url }%
\providecommand \@url [1]{\endgroup\@href {#1}{\urlprefix }}%
\providecommand \urlprefix  [0]{URL }%
\providecommand \Eprint [0]{\href }%
\providecommand \doibase [0]{http://dx.doi.org/}%
\providecommand \selectlanguage [0]{\@gobble}%
\providecommand \bibinfo  [0]{\@secondoftwo}%
\providecommand \bibfield  [0]{\@secondoftwo}%
\providecommand \translation [1]{[#1]}%
\providecommand \BibitemOpen [0]{}%
\providecommand \bibitemStop [0]{}%
\providecommand \bibitemNoStop [0]{.\EOS\space}%
\providecommand \EOS [0]{\spacefactor3000\relax}%
\providecommand \BibitemShut  [1]{\csname bibitem#1\endcsname}%
\let\auto@bib@innerbib\@empty
%</preamble>
\bibitem [{\citenamefont {Wiener}(1956)}]{Wiener_1956}%
  \BibitemOpen
  \bibfield  {author} {\bibinfo {author} {\bibfnamefont {N.}~\bibnamefont
  {Wiener}},\ }\href@noop {} {\emph {\bibinfo {title} {The theory of prediction
  in Modern Mathematics for the Engineer}}}\ (\bibinfo  {publisher} {New York:
  McGraw-Hill},\ \bibinfo {year} {1956})\ pp.\ \bibinfo {pages}
  {165--190}\BibitemShut {NoStop}%
\bibitem [{\citenamefont {Granger}(1969)}]{Granger_1969}%
  \BibitemOpen
  \bibfield  {author} {\bibinfo {author} {\bibfnamefont {C.~W.~J.}\
  \bibnamefont {Granger}},\ }\href {\doibase 10.2307/1912791} {\bibfield
  {journal} {\bibinfo  {journal} {Econometrica}\ }\textbf {\bibinfo {volume}
  {37}},\ \bibinfo {pages} {424} (\bibinfo {year} {1969})}\BibitemShut
  {NoStop}%
\bibitem [{\citenamefont {Granger}(1980)}]{Granger_1980}%
  \BibitemOpen
  \bibfield  {author} {\bibinfo {author} {\bibfnamefont {C.}~\bibnamefont
  {Granger}},\ }\href {\doibase https://doi.org/10.1016/0165-1889(80)90069-X}
  {\bibfield  {journal} {\bibinfo  {journal} {Journal of Economic Dynamics and
  Control}\ }\textbf {\bibinfo {volume} {2}},\ \bibinfo {pages} {329 }
  (\bibinfo {year} {1980})}\BibitemShut {NoStop}%
\bibitem [{\citenamefont {Baccala}\ \emph {et~al.}(1998)\citenamefont
  {Baccala}, \citenamefont {Sameshima}, \citenamefont {Ballester},\ and\
  \citenamefont {do~Valle}}]{Sameshima_1998}%
  \BibitemOpen
  \bibfield  {author} {\bibinfo {author} {\bibfnamefont {L.}~\bibnamefont
  {Baccala}}, \bibinfo {author} {\bibfnamefont {K.}~\bibnamefont {Sameshima}},
  \bibinfo {author} {\bibfnamefont {G.}~\bibnamefont {Ballester}}, \ and\
  \bibinfo {author} {\bibfnamefont {A.~C.}\ \bibnamefont {do~Valle}},\
  }\href@noop {} {\bibfield  {journal} {\bibinfo  {journal} {Applied Signal
  Processing}\ }\textbf {\bibinfo {volume} {5}} (\bibinfo {year}
  {1998})}\BibitemShut {NoStop}%
\bibitem [{\citenamefont {Gourevitch}\ \emph {et~al.}(2006)\citenamefont
  {Gourevitch}, \citenamefont {Jeannes},\ and\ \citenamefont
  {Faucon}}]{Gourevitch_2006}%
  \BibitemOpen
  \bibfield  {author} {\bibinfo {author} {\bibfnamefont {B.}~\bibnamefont
  {Gourevitch}}, \bibinfo {author} {\bibfnamefont {R.~L.~B.}\ \bibnamefont
  {Jeannes}}, \ and\ \bibinfo {author} {\bibfnamefont {G.}~\bibnamefont
  {Faucon}},\ }\href@noop {} {\bibfield  {journal} {\bibinfo  {journal}
  {Biological Cybernetics}\ }\textbf {\bibinfo {volume} {95}} (\bibinfo {year}
  {2006})}\BibitemShut {NoStop}%
\bibitem [{\citenamefont {Barrett}\ \emph {et~al.}(2015)\citenamefont
  {Barrett}, \citenamefont {Murphy}, \citenamefont {Bruno}, \citenamefont
  {Norhomme}, \citenamefont {Boly}, \citenamefont {Laureys},\ and\
  \citenamefont {Seth}}]{EEG_2012}%
  \BibitemOpen
  \bibfield  {author} {\bibinfo {author} {\bibfnamefont {A.~B.}\ \bibnamefont
  {Barrett}}, \bibinfo {author} {\bibfnamefont {M.}~\bibnamefont {Murphy}},
  \bibinfo {author} {\bibfnamefont {M.-A.}\ \bibnamefont {Bruno}}, \bibinfo
  {author} {\bibfnamefont {Q.}~\bibnamefont {Norhomme}}, \bibinfo {author}
  {\bibfnamefont {M.}~\bibnamefont {Boly}}, \bibinfo {author} {\bibfnamefont
  {S.}~\bibnamefont {Laureys}}, \ and\ \bibinfo {author} {\bibfnamefont
  {A.~K.}\ \bibnamefont {Seth}},\ }\href {\doibase
  10.1523/JNEUROSCI.4399-14.2015} {\bibfield  {journal} {\bibinfo  {journal}
  {Journal of Neuroscience}\ }\textbf {\bibinfo {volume} {35}},\ \bibinfo
  {pages} {3293} (\bibinfo {year} {2015})}\BibitemShut {NoStop}%
\bibitem [{\citenamefont {Ding}\ \emph {et~al.}(2006)\citenamefont {Ding},
  \citenamefont {Chen},\ and\ \citenamefont {Bressler}}]{appl_book}%
  \BibitemOpen
  \bibfield  {author} {\bibinfo {author} {\bibfnamefont {M.}~\bibnamefont
  {Ding}}, \bibinfo {author} {\bibfnamefont {Y.}~\bibnamefont {Chen}}, \ and\
  \bibinfo {author} {\bibfnamefont {S.~L.}\ \bibnamefont {Bressler}},\ }\href
  {\doibase 10.1002/9783527609970.ch17} {\emph {\bibinfo {title} {Granger
  Causality: Basic Theory and Applications to Neuroscience. In: Schlter S,
  Winterhalder N., Timmer J., eds. handbook of time series analysis}}}\
  (\bibinfo  {publisher} {Wienheim: Wiley},\ \bibinfo {year} {2006})\
  Chap.~\bibinfo {chapter} {17}, pp.\ \bibinfo {pages} {437--460},\ \Eprint
  {http://arxiv.org/abs/https://onlinelibrary.wiley.com/doi/pdf/10.1002/9783527609970.ch17}
  {https://onlinelibrary.wiley.com/doi/pdf/10.1002/9783527609970.ch17}
  \BibitemShut {NoStop}%
\bibitem [{\citenamefont {Chen}\ \emph {et~al.}(2006)\citenamefont {Chen},
  \citenamefont {Bressler},\ and\ \citenamefont {Ding}}]{Chen_2006}%
  \BibitemOpen
  \bibfield  {author} {\bibinfo {author} {\bibfnamefont {Y.}~\bibnamefont
  {Chen}}, \bibinfo {author} {\bibfnamefont {S.~L.}\ \bibnamefont {Bressler}},
  \ and\ \bibinfo {author} {\bibfnamefont {M.}~\bibnamefont {Ding}},\ }\href
  {\doibase https://doi.org/10.1016/j.jneumeth.2005.06.011} {\bibfield
  {journal} {\bibinfo  {journal} {Journal of Neuroscience Methods}\ }\textbf
  {\bibinfo {volume} {150}},\ \bibinfo {pages} {228 } (\bibinfo {year}
  {2006})}\BibitemShut {NoStop}%
\bibitem [{\citenamefont {Bressler}\ and\ \citenamefont {Seth}(2011)}]{wem}%
  \BibitemOpen
  \bibfield  {author} {\bibinfo {author} {\bibfnamefont {S.~L.}\ \bibnamefont
  {Bressler}}\ and\ \bibinfo {author} {\bibfnamefont {A.~K.}\ \bibnamefont
  {Seth}},\ }\href {\doibase https://doi.org/10.1016/j.neuroimage.2010.02.059}
  {\bibfield  {journal} {\bibinfo  {journal} {NeuroImage}\ }\textbf {\bibinfo
  {volume} {58}},\ \bibinfo {pages} {323 } (\bibinfo {year}
  {2011})}\BibitemShut {NoStop}%
\bibitem [{\citenamefont {Seth}\ \emph {et~al.}(2015)\citenamefont {Seth},
  \citenamefont {Barrett},\ and\ \citenamefont {Barnett}}]{Seth_jns}%
  \BibitemOpen
  \bibfield  {author} {\bibinfo {author} {\bibfnamefont {A.~K.}\ \bibnamefont
  {Seth}}, \bibinfo {author} {\bibfnamefont {A.~B.}\ \bibnamefont {Barrett}}, \
  and\ \bibinfo {author} {\bibfnamefont {L.}~\bibnamefont {Barnett}},\ }\href
  {\doibase 10.1523/JNEUROSCI.4399-14.2015} {\bibfield  {journal} {\bibinfo
  {journal} {Journal of Neuroscience}\ }\textbf {\bibinfo {volume} {35}},\
  \bibinfo {pages} {3293} (\bibinfo {year} {2015})}\BibitemShut {NoStop}%
\bibitem [{\citenamefont {{De Vico Fallani}}\ \emph {et~al.}(2015)\citenamefont
  {{De Vico Fallani}}, \citenamefont {{Corazzol}}, \citenamefont {{Sternberg}},
  \citenamefont {{Wyart}},\ and\ \citenamefont {{Chavez}}}]{CID_2015}%
  \BibitemOpen
  \bibfield  {author} {\bibinfo {author} {\bibfnamefont {F.}~\bibnamefont {{De
  Vico Fallani}}}, \bibinfo {author} {\bibfnamefont {M.}~\bibnamefont
  {{Corazzol}}}, \bibinfo {author} {\bibfnamefont {J.~R.}\ \bibnamefont
  {{Sternberg}}}, \bibinfo {author} {\bibfnamefont {C.}~\bibnamefont
  {{Wyart}}}, \ and\ \bibinfo {author} {\bibfnamefont {M.}~\bibnamefont
  {{Chavez}}},\ }\href {\doibase 10.1109/TNSRE.2014.2341632} {\bibfield
  {journal} {\bibinfo  {journal} {IEEE Transactions on Neural Systems and
  Rehabilitation Engineering}\ }\textbf {\bibinfo {volume} {23}},\ \bibinfo
  {pages} {333} (\bibinfo {year} {2015})}\BibitemShut {NoStop}%
\bibitem [{\citenamefont {Stevenson}\ \emph {et~al.}(2009)\citenamefont
  {Stevenson}, \citenamefont {Rebesco}, \citenamefont {Hatsopoulos},
  \citenamefont {Haga}, \citenamefont {Miller},\ and\ \citenamefont
  {Kording}}]{Stevenson_2009}%
  \BibitemOpen
  \bibfield  {author} {\bibinfo {author} {\bibfnamefont {H.~I.}\ \bibnamefont
  {Stevenson}}, \bibinfo {author} {\bibfnamefont {M.~J.}\ \bibnamefont
  {Rebesco}}, \bibinfo {author} {\bibfnamefont {G.~N.}\ \bibnamefont
  {Hatsopoulos}}, \bibinfo {author} {\bibfnamefont {Z.}~\bibnamefont {Haga}},
  \bibinfo {author} {\bibfnamefont {E.~L.}\ \bibnamefont {Miller}}, \ and\
  \bibinfo {author} {\bibfnamefont {P.~K.}\ \bibnamefont {Kording}},\
  }\href@noop {} {\bibfield  {journal} {\bibinfo  {journal} {IEEE Transactions
  on Neural Systems and Rehabilitation Engineering}\ }\textbf {\bibinfo
  {volume} {17}},\ \bibinfo {pages} {203} (\bibinfo {year} {2009})}\BibitemShut
  {NoStop}%
\bibitem [{\citenamefont {Kim}\ \emph {et~al.}(2011)\citenamefont {Kim},
  \citenamefont {Putrino}, \citenamefont {Ghosh},\ and\ \citenamefont
  {Brown}}]{Kim_2011}%
  \BibitemOpen
  \bibfield  {author} {\bibinfo {author} {\bibfnamefont {S.}~\bibnamefont
  {Kim}}, \bibinfo {author} {\bibfnamefont {D.}~\bibnamefont {Putrino}},
  \bibinfo {author} {\bibfnamefont {S.}~\bibnamefont {Ghosh}}, \ and\ \bibinfo
  {author} {\bibfnamefont {N.~E.}\ \bibnamefont {Brown}},\ }\href@noop {}
  {\bibfield  {journal} {\bibinfo  {journal} {Plos Computational Biology}\
  }\textbf {\bibinfo {volume} {7}},\ \bibinfo {pages} {e1001110} (\bibinfo
  {year} {2011})}\BibitemShut {NoStop}%
\bibitem [{\citenamefont {Gerhard}\ \emph {et~al.}(2013)\citenamefont
  {Gerhard}, \citenamefont {Kispersky}, \citenamefont {Gutierrez},
  \citenamefont {Marder}, \citenamefont {Kramer},\ and\ \citenamefont
  {Eden}}]{Eden_2013}%
  \BibitemOpen
  \bibfield  {author} {\bibinfo {author} {\bibfnamefont {F.}~\bibnamefont
  {Gerhard}}, \bibinfo {author} {\bibfnamefont {T.}~\bibnamefont {Kispersky}},
  \bibinfo {author} {\bibfnamefont {G.~J.}\ \bibnamefont {Gutierrez}}, \bibinfo
  {author} {\bibfnamefont {E.}~\bibnamefont {Marder}}, \bibinfo {author}
  {\bibfnamefont {M.}~\bibnamefont {Kramer}}, \ and\ \bibinfo {author}
  {\bibfnamefont {U.}~\bibnamefont {Eden}},\ }\href {\doibase
  10.1371/journal.pcbi.1003138} {\bibfield  {journal} {\bibinfo  {journal}
  {Plos Computational Biology}\ }\textbf {\bibinfo {volume} {9}},\ \bibinfo
  {pages} {e1003138} (\bibinfo {year} {2013})}\BibitemShut {NoStop}%
\bibitem [{\citenamefont {Roebroeck}\ \emph {et~al.}(2011)\citenamefont
  {Roebroeck}, \citenamefont {Formisano},\ and\ \citenamefont
  {Goebel}}]{brain_selection}%
  \BibitemOpen
  \bibfield  {author} {\bibinfo {author} {\bibfnamefont {A.}~\bibnamefont
  {Roebroeck}}, \bibinfo {author} {\bibfnamefont {E.}~\bibnamefont
  {Formisano}}, \ and\ \bibinfo {author} {\bibfnamefont {R.}~\bibnamefont
  {Goebel}},\ }\href {\doibase
  https://doi.org/10.1016/j.neuroimage.2009.09.036} {\bibfield  {journal}
  {\bibinfo  {journal} {NeuroImage}\ }\textbf {\bibinfo {volume} {58}},\
  \bibinfo {pages} {296 } (\bibinfo {year} {2011})}\BibitemShut {NoStop}%
\bibitem [{\citenamefont {Pearl}(2000)}]{Pearl_2000}%
  \BibitemOpen
  \bibfield  {author} {\bibinfo {author} {\bibfnamefont {J.}~\bibnamefont
  {Pearl}},\ }\href@noop {} {\emph {\bibinfo {title} {Causality: Models,
  Reasoning, and Inference}}}\ (\bibinfo  {publisher} {Cambridge University
  Press},\ \bibinfo {year} {2000})\BibitemShut {NoStop}%
\bibitem [{\citenamefont {Guo}\ \emph {et~al.}(2008)\citenamefont {Guo},
  \citenamefont {Seth}, \citenamefont {Kendrick}, \citenamefont {Zhou},\ and\
  \citenamefont {Feng}}]{Guo_2008}%
  \BibitemOpen
  \bibfield  {author} {\bibinfo {author} {\bibfnamefont {S.}~\bibnamefont
  {Guo}}, \bibinfo {author} {\bibfnamefont {A.~K.}\ \bibnamefont {Seth}},
  \bibinfo {author} {\bibfnamefont {K.~M.}\ \bibnamefont {Kendrick}}, \bibinfo
  {author} {\bibfnamefont {C.}~\bibnamefont {Zhou}}, \ and\ \bibinfo {author}
  {\bibfnamefont {J.}~\bibnamefont {Feng}},\ }\href {\doibase
  https://doi.org/10.1016/j.jneumeth.2008.04.011} {\bibfield  {journal}
  {\bibinfo  {journal} {Journal of Neuroscience Methods}\ }\textbf {\bibinfo
  {volume} {172}},\ \bibinfo {pages} {79 } (\bibinfo {year}
  {2008})}\BibitemShut {NoStop}%
\bibitem [{\citenamefont {Smith}\ \emph {et~al.}(2011)\citenamefont {Smith},
  \citenamefont {Miller}, \citenamefont {Salimi-Khorshidi}, \citenamefont
  {Webster}, \citenamefont {Beckmann}, \citenamefont {Nichols}, \citenamefont
  {Ramsey},\ and\ \citenamefont {Woolrich}}]{Smith_2011}%
  \BibitemOpen
  \bibfield  {author} {\bibinfo {author} {\bibfnamefont {S.~M.}\ \bibnamefont
  {Smith}}, \bibinfo {author} {\bibfnamefont {K.~L.}\ \bibnamefont {Miller}},
  \bibinfo {author} {\bibfnamefont {G.}~\bibnamefont {Salimi-Khorshidi}},
  \bibinfo {author} {\bibfnamefont {M.}~\bibnamefont {Webster}}, \bibinfo
  {author} {\bibfnamefont {C.~F.}\ \bibnamefont {Beckmann}}, \bibinfo {author}
  {\bibfnamefont {T.~E.}\ \bibnamefont {Nichols}}, \bibinfo {author}
  {\bibfnamefont {J.~D.}\ \bibnamefont {Ramsey}}, \ and\ \bibinfo {author}
  {\bibfnamefont {M.~W.}\ \bibnamefont {Woolrich}},\ }\href {\doibase
  https://doi.org/10.1016/j.neuroimage.2010.08.063} {\bibfield  {journal}
  {\bibinfo  {journal} {NeuroImage}\ }\textbf {\bibinfo {volume} {54}},\
  \bibinfo {pages} {875 } (\bibinfo {year} {2011})}\BibitemShut {NoStop}%
\bibitem [{\citenamefont {Roelstraete}\ and\ \citenamefont
  {Rosseel}(2012)}]{dpgc}%
  \BibitemOpen
  \bibfield  {author} {\bibinfo {author} {\bibfnamefont {B.}~\bibnamefont
  {Roelstraete}}\ and\ \bibinfo {author} {\bibfnamefont {Y.}~\bibnamefont
  {Rosseel}},\ }\href {\doibase https://doi.org/10.1016/j.jneumeth.2012.01.010}
  {\bibfield  {journal} {\bibinfo  {journal} {Journal of Neuroscience Methods}\
  }\textbf {\bibinfo {volume} {206}},\ \bibinfo {pages} {73 } (\bibinfo {year}
  {2012})}\BibitemShut {NoStop}%
\bibitem [{\citenamefont {Barrett}\ \emph {et~al.}(2010)\citenamefont
  {Barrett}, \citenamefont {Barnett},\ and\ \citenamefont
  {Seth}}]{Barrett_2010}%
  \BibitemOpen
  \bibfield  {author} {\bibinfo {author} {\bibfnamefont {A.~B.}\ \bibnamefont
  {Barrett}}, \bibinfo {author} {\bibfnamefont {L.}~\bibnamefont {Barnett}}, \
  and\ \bibinfo {author} {\bibfnamefont {A.~K.}\ \bibnamefont {Seth}},\ }\href
  {\doibase 10.1103/PhysRevE.81.041907} {\bibfield  {journal} {\bibinfo
  {journal} {Phys. Rev. E}\ }\textbf {\bibinfo {volume} {81}},\ \bibinfo
  {pages} {041907} (\bibinfo {year} {2010})}\BibitemShut {NoStop}%
\bibitem [{\citenamefont {Barnett}\ and\ \citenamefont {Seth}(2015)}]{ssgc}%
  \BibitemOpen
  \bibfield  {author} {\bibinfo {author} {\bibfnamefont {L.}~\bibnamefont
  {Barnett}}\ and\ \bibinfo {author} {\bibfnamefont {A.~K.}\ \bibnamefont
  {Seth}},\ }\href {\doibase 10.1103/PhysRevE.91.040101} {\bibfield  {journal}
  {\bibinfo  {journal} {Phys. Rev. E}\ }\textbf {\bibinfo {volume} {91}},\
  \bibinfo {pages} {040101} (\bibinfo {year} {2015})}\BibitemShut {NoStop}%
\bibitem [{\citenamefont {Dobson}\ and\ \citenamefont
  {Barnett}(2008)}]{Dobson}%
  \BibitemOpen
  \bibfield  {author} {\bibinfo {author} {\bibfnamefont {A.~J.}\ \bibnamefont
  {Dobson}}\ and\ \bibinfo {author} {\bibfnamefont {A.~G.}\ \bibnamefont
  {Barnett}},\ }\href@noop {} {\emph {\bibinfo {title} {An Introduction to
  Generalized Linear Models, Third Edition}}}\ (\bibinfo  {publisher} {CRC
  Press},\ \bibinfo {year} {2008})\BibitemShut {NoStop}%
\bibitem [{\citenamefont {Mccullagh}\ and\ \citenamefont
  {Nelder}(2008)}]{Mccullagh}%
  \BibitemOpen
  \bibfield  {author} {\bibinfo {author} {\bibfnamefont {P.}~\bibnamefont
  {Mccullagh}}\ and\ \bibinfo {author} {\bibfnamefont {J.~A.}\ \bibnamefont
  {Nelder}},\ }\href@noop {} {\emph {\bibinfo {title} {Generalized Linear
  Models, Second Edition}}}\ (\bibinfo  {publisher} {Chapman and Hall},\
  \bibinfo {year} {2008})\BibitemShut {NoStop}%
\bibitem [{\citenamefont {Baccal{\'a}}\ and\ \citenamefont
  {Sameshima}(2001)}]{Baccal_2001}%
  \BibitemOpen
  \bibfield  {author} {\bibinfo {author} {\bibfnamefont {L.~A.}\ \bibnamefont
  {Baccal{\'a}}}\ and\ \bibinfo {author} {\bibfnamefont {K.}~\bibnamefont
  {Sameshima}},\ }\href {\doibase 10.1007/PL00007990} {\bibfield  {journal}
  {\bibinfo  {journal} {Biological Cybernetics}\ }\textbf {\bibinfo {volume}
  {84}},\ \bibinfo {pages} {463} (\bibinfo {year} {2001})}\BibitemShut
  {NoStop}%
\end{thebibliography}%

\end{document}